\documentclass[10pt,a4paper,english,prd,nofootinbib,notitlepage,superscriptaddress,showkeys,preprintnumbers,longbibliography]{revtex4-1}
\usepackage[T1]{fontenc}
\usepackage[latin9]{inputenc}
\usepackage{color}
\usepackage{babel}
\usepackage{amsmath}
\usepackage{amssymb}
\usepackage{graphicx}
\usepackage{esint}
\usepackage[unicode=true,
 bookmarks=false,
 breaklinks=false,pdfborder={0 0 1},backref=false,colorlinks=true]
 {hyperref}
\hypersetup{
 citecolor=blue,urlcolor=blue,linkcolor=blue}
\usepackage{breakurl}

\makeatletter

\providecommand{\tabularnewline}{\\}

\usepackage{babel}

\usepackage{babel}
\usepackage{breakurl}\usepackage{cleveref}\usepackage{ulem}\usepackage{bbm}

\usepackage{todonotes}
\definecolor{lightblue}{HTML}{A9D0F5}
\definecolor{lightgreen}{HTML}{BCF5A9}
\definecolor{lightred}{HTML}{F6CECE}
\definecolor{lightorange}{HTML}{FFA800}
\definecolor{greengray}{HTML}{5C9393}
\definecolor{lightgreengray}{HTML}{80CCCC}
\providecommand{\tabularnewline}{\\}

\newcommand{\Mp}{M_\mathrm{Pl}}

\makeatother

\begin{document}

\title{Dynamical analysis of $R\dfrac{1}{\Box^{2}}R$ cosmology: \\
 Impact of initial conditions and constraints from supernovae}

\author{Henrik Nersisyan}

\email{h.nersisyan@thphys.uni-heidelberg.de}

\affiliation{Institut f\"{u}r Theoretische Physik, Ruprecht-Karls-Universit\"{a}t Heidelberg,
Philosophenweg 16, 69120 Heidelberg, Germany}

\author{Yashar Akrami}

\email{y.akrami@thphys.uni-heidelberg.de}

\affiliation{Institut f\"{u}r Theoretische Physik, Ruprecht-Karls-Universit\"{a}t Heidelberg,
Philosophenweg 16, 69120 Heidelberg, Germany}

\author{Luca Amendola}

\email{l.amendola@thphys.uni-heidelberg.de}

\affiliation{Institut f\"{u}r Theoretische Physik, Ruprecht-Karls-Universit\"{a}t Heidelberg,
Philosophenweg 16, 69120 Heidelberg, Germany}

\author{Tomi S. Koivisto}

\email{tomi.koivisto@nordita.org}

\preprint{NORDITA-2016-39}

\affiliation{Nordita, KTH Royal Institute of Technology and Stockholm University,
Roslagstullsbacken 23, 10691 Stockholm, Sweden}

\author{Javier Rubio}

\email{j.rubio@thphys.uni-heidelberg.de}

\affiliation{Institut f\"{u}r Theoretische Physik, Ruprecht-Karls-Universit\"{a}t Heidelberg,
Philosophenweg 16, 69120 Heidelberg, Germany}
\begin{abstract}
We discuss the cosmological implications of the $R~\Box^{-2}R$ nonlocal
modification to standard gravity. We relax the assumption of special
initial conditions in the local formulation of the theory, perform
a full phase-space analysis of the system, and show that the late-time
cosmology of the model exhibits two distinct evolution paths, on which a large range of values for the present equation of state can be reached. We then
compare the general solutions to supernovae data and place constraints
on the parameters of the model. In particular, we find that the mass parameter of the theory should be smaller than 1.2 in Hubble units.

\end{abstract}


\date{\today}
\maketitle

\section{Introduction}

The current standard model of cosmology, called $\Lambda$CDM (CDM
for cold dark matter), cannot be reconciled with general relativity
(GR) and the Standard Model of particle physics without extreme fine-tuning.
In particular, the ratio $\sqrt{\Lambda}/\Mp^{2}$ derived from observations
[with $\Lambda$ the notorious cosmological constant (CC) and $\Mp$
the reduced Planck mass ] is almost infinitesimal compared to the value
obtained by the most straightforward extrapolations of GR and quantum
field theory, to the infrared scale $\sqrt{\Lambda}/\Mp$ and high-energy
scales approaching $\Mp$, respectively. This calls both for the observational
pursuit of signatures that could provide hints on the possible physics
beyond the $\Lambda$CDM model, and for theoretical extensions that could
explain the cosmological data in a more natural way \cite{2010deto.book.....A,Bull:2015stt}.

Various attempts at such extensions have been undertaken in the context
of nonlocal gravity \cite{Woodard:2014iga,Barvinsky:2014lja}. In
a top-bottom approach, the possibility that gravitational interactions
become nonlocal near the Planck scale is suggested, among others,
by string theory \cite{Siegel:2003vt,Calcagni:2014vxa}. From a bottom-up
perspective, nonlocal theories are appealing because of their potential
to provide an ultraviolet completion of the metric gravity theory
\cite{Biswas:2013kla,Biswas:2011ar,Modesto:2011kw}, but there are
also motivations to contemplate nonlocal terms in the infrared as
well. Such infrared nonlocal terms arise generically in effective
field theories after integrating out light degrees of freedom \cite{Barvinsky:2014lja,Codello:2015mba,Donoghue:2015nba},
but may also feature in more fundamental actions in Euclidean quantum
gravity \cite{Wetterich:1997bz,Barvinsky:2011rk}. Nonlocal effective
formulations have been found for gravity models with a massive graviton
\cite{Jaccard:2013gla,Modesto:2013jea}, multiple metrics \cite{Cusin:2014zoa},
and post-Riemannian, affine geometry \cite{Golovnev:2015bsa}. In
passing, we note that indeed the recent development of a conformal
affine gauge theory of gravity \cite{Frank} introduces a novel holography
that, along the lines of Ref.~\cite{Padmanabhan:2016oml}, may naturally
provide a nonlocal link between the value of the cosmological constant
and the amount of information contained in the emergent spacetime.

Nonlocal gravity models are typically written as an Einstein-Hilbert
term supplemented with integral or infinite-derivative curvature terms.
The first proposal for a nonlocal dark-energy model was put forward
by Deser and Woodard (DW) and has the form \cite{Deser:2007jk} 
\begin{equation}
\mathcal{L}^{\text{DW}}=\frac{\Mp^{2}}{2}R\left[1-f\left(\frac{R}{\Box}\right)\right],\label{DWaction}
\end{equation}
where $R$ is the Ricci curvature scalar and $1/\Box$ is the inverse
d'Alembertian, an integral operator such that $\Box(1/\Box)=1$, with
$\Box\equiv g^{\mu\nu}\nabla_{\mu}\nabla_{\nu}$ and $\nabla_{\mu}$
the Christoffel covariant derivative. With the dimensionless combination
$R/\Box$, one could in principle construct models without introducing
new scales. The integral dependence of the corrections could generate
the observed acceleration at the present cosmological epoch dynamically
and without special fine-tunings. However, detailed investigations
have shown that, although the function $f$ can be chosen in such
a way that the background expansion is consistent with the data \cite{Koivisto:2008xfa,Deffayet:2009ca,Elizalde:2012ja}
and the model has a viable Newtonian limit \cite{Koivisto:2008dh,Conroy:2014eja},
the impact of the nonlocal corrections on the evolution of perturbations
is strong and utterly rules the model out when this is confronted
with large-scale structure data \cite{Dodelson:2013sma}. On top of
that, nonlocal modifications of gravity result generically in instabilities
at the level of perturbations, at least if they involve tensorial
terms such as $(W_{\mu\nu\rho\sigma}/\Box^{2})W^{\mu\nu\rho\sigma}$
\cite{Ferreira:2013tqn} with $W_{\mu\nu\rho\sigma}$ the Weyl tensor
appearing in models inspired by the conformal anomaly \cite{Cusin:2015rex,Cusin:2016nzi}.

One of the remarkable features of the model 
\begin{equation}
\mathcal{L}^{\text{MM}}=\frac{\Mp^{2}}{2}R\left[1-\frac{m^{2}}{6}\left(\frac{1}{\Box}\right)^{2}R\right]=\frac{\Mp^{2}}{2}\left[R-\frac{m^{2}}{6}\left(\frac{R}{\Box}\right)^{2}\right],\label{NLaction1}
\end{equation}
proposed by Maggiore and Mancarella (MM) \cite{Maggiore:2014sia}
is that it can produce nonlocal dark energy able to fit the background
data while retaining a matter power spectrum compatible with observations
(see Refs.~\cite{Dirian:2014xoa,Dirian:2014bma,Codello:2015pga,Codello:2016neo}
and \cite{Dirian:2014ara,Nesseris:2014mea,Barreira:2014kra,Dirian:2016puz}
for studies of the background expansion and of structure formation,
respectively). It is also notable that the $(R/\Box)^{2}$-correction
to GR has indeed been obtained in an effective field theory for gravity
at the second order curvature expansion%
\footnote{As shown in Ref.~\cite{Maggiore:2016fbn}, the coefficient of the $R\Box^{-2}R$ obtained by
this procedure should satisfy $M^2/H^2\ll 1$ with  $M^4\sim (M_{Pl}\, m)^2$. Unfortunately, this condition is 
not compatible with the value of $m$ required to obtain a realistic cosmology ($m\sim H_0$).} \cite{Codello:2015mba} and that the MM model appears to have only
one new parameter $m$ at the level of the gravitational Lagrangian,
i.e. none more than $\Lambda$CDM.%
\footnote{Expectedly, viable dark energy models require $m\sim\Lambda/\Mp\sim H_{0}$,
where $H_{0}$ is the present Hubble rate.%
} It has also been argued that ghost fields do not destabilize the
model \cite{Maggiore:2014sia} (see also Ref.~\cite{Foffa:2013vma}).
Spherically symmetric solutions have also been considered \cite{Kehagias:2014sda,Conroy:2014eja}.

In this paper, we study the cosmological dynamics of the MM model,
with special attention to the problem of initial conditions. Nonlocal
theories with infinite order derivative operators require the specification
of an infinite number of initial conditions for the formulation of
the Cauchy problem. Analogously, nonlocal integral operators, such
as the one featured in the MM model, are strictly defined only by
specifying the boundary conditions for each of the infinite number
of modes in the continuum limit of the Fourier space. Various techniques
have been considered to deal with such theories, see Refs.~\cite{Woodard:2000bt,Barnaby:2008tc,Koivisto:2009jn,Barnaby:2010kx,Calcagni:2010ab,Foffa:2013sma,Tsamis:2014hra,Zhang:2016ykx}.
The MM model (\ref{NLaction1}) can be reformulated in terms of two
scalar fields \cite{Koivisto:2008dh}, which should not be considered
however as local dynamical fields evolving freely in time, but as
auxiliary fields whose configuration at each spatial hypersurface
is dictated by the other fields and the boundary conditions of the
$1/\Box$-operator. In the phase space of the homogeneous cosmological
dynamics, the trajectories of the two (fake) scalar degrees of freedom
are uniquely fixed given four numbers at any given cosmological epoch.
The cosmology of the MM model seems to offer a natural or ``minimal''
assumption for the choice of these numbers: at a sufficiently early
epoch in the standard cosmology, the Universe is filled with radiation
only, for which $R\approx0$. It therefore seems an obvious choice
to set $R/\Box=R/\Box^{2}=0$ at such an epoch.%
\footnote{Note however that the cosmology obviously depends on the
thermal history. In Appendix \ref{appendix2}, we check the impact
of setting $R/\Box=R/\Box^{2}=0$ either at the matter-radiation equality
or at an earlier period.%
} However, already at the linear order in the inhomogeneous fluctuations,
both the inverse- and the double-inverse-d'Alembertian operators bring
forth scale-dependent functions in the momentum space. Unless finely adjusted and compensating scale dependence is encoded into the boundary
conditions of the $1/\Box$-operators, the initial conditions for
cosmological perturbations would feature additional scale dependence
(compared to $\Lambda$CDM). The minimal boundary conditions,
that is $\delta(R/\Box)=0$ when $\delta R=0$ (we denote perturbations
with $\delta$), would require scale dependence in the initial conditions
for the auxiliary fields. An important point is that due to their
assumed nonlocal origin, they impose constraints rather than adding dynamics.
Thus one expects the nonminimal scale dependence of the initial conditions
to be directly projected (or, if set in terms of the auxiliary fields,
to effectively propagate) to the smaller redshifts of the crucial
observables, where especially the matter power spectrum is very sensitive
to the possible scale dependence in the dark sector, as that is reflected
through the gravitational interaction in the baryon distribution.
Since the confrontation with large-scale structure is crucial for
distinguishing the MM (\ref{NLaction1}) and the earlier proposal
(\ref{DWaction}), the issue of (scale-dependent) linear boundary
conditions calls for clarification.

In this paper we undertake a comprehensive study of the expansion
dynamics in the MM model. In Sec.~\ref{sec:cosmology} we rewrite
the model (\ref{NLaction1}) in terms of two (effective) auxiliary
scalar fields, and set up the phase space spanned by convenient dimensionless
variables whose dynamical system can be closed into an autonomous
form. In Sec.~\ref{sec:dynanalysis} we perform a full dynamical
system analysis in order to identify the critical points in the cosmological
phase space and determine their stability. Each set of initial conditions
fixes a trajectory in the phase space, corresponding to a particular
family of MM models with the same mass parameter $m$ and the same
four cosmological background boundary conditions. By exploring the
global structure of the phase space we can thus map the cosmology
of different models and investigate the sensitivity of the predictions
to changing the parameters of the model (i.e. to the initial conditions
that have been previously assumed minimal). In Sec.~\ref{sec:SNe}
we confront the model with supernovae data constraining the background
expansion, in such a way that we do not fix all the initial conditions
but marginalize over them. Our findings are then summarized in Sec.~\ref{sec:conclusions}.

\section{The cosmology of $R\dfrac{1}{\Box^{2}}R$ gravity model}

\label{sec:cosmology}

The full action, including both gravity and matter sectors, for the
MM nonlocal theory introduced in Eq. (\ref{NLaction1}) has the form
\begin{equation}
S^{\text{MM}}=\frac{\Mp^{2}}{2}\int\text{d}^{4}x\sqrt{-g}\left(R-\frac{m^{2}}{6}R\dfrac{1}{\Box^{2}}R\right)+\int\text{d}^{4}x\sqrt{-g}\mathfrak{\mathit{\mathcal{L}_{m}}},\label{NLaction}
\end{equation}
with the mass scale $m$ the only free parameter of the theory, to
be determined observationally, and $\mathfrak{\mathit{\mathcal{L}_{m}}}$
the matter Lagrangian minimally coupled to gravity.

In order to derive the modified Einstein equations, we vary the action~(\ref{NLaction})
with respect to the metric $g_{\mu\nu}$: 
\begin{align}
\delta S^{\text{MM}} & =\frac{\Mp^{2}}{2}\int\text{d}^{4}x\delta\left(\sqrt{-g}\right)\left(R-\frac{m^{2}}{6}R\dfrac{1}{\Box^{2}}R\right)\\
 & +\frac{\Mp^{2}}{2}\int\text{d}^{4}x\sqrt{-g}\left(\delta R-\frac{m^{2}}{3}\delta R\dfrac{1}{\Box^{2}}R+\frac{m^{2}}{3}R\dfrac{1}{\Box}\delta\Box\left(\dfrac{1}{\Box^{2}}R\right)\right)+\delta\int\text{d}^{4}x\sqrt{-g}\mathfrak{\mathit{\mathcal{L}_{m}}},\nonumber 
\end{align}
where we have used $\delta(\Box^{-2})=-2\Box^{-1}(\delta\Box)\Box^{-2}$.
Denoting the conserved stress-energy tensor of matter by $T_{\nu}^{\mu}$,
the gravitational field equations turn out to be~\cite{Maggiore:2014sia}
\begin{equation}
G_{\nu}^{\mu}-\frac{1}{6}m^{2}K_{\nu}^{\mu}=8\pi GT_{\nu}^{\mu},\label{eqmRiBox2R}
\end{equation}
where we have defined 
\begin{align}
K_{\nu}^{\mu} & \equiv2SG_{\nu}^{\mu}-2\nabla^{\mu}\partial_{\nu}S+2\delta_{\nu}^{\mu}\Box S+\delta_{\nu}^{\mu}\partial_{\rho}S\partial^{\rho}U-\frac{1}{2}\delta_{\nu}^{\mu}U^{2}-\big(\partial^{\mu}S\partial_{\nu}U+\partial_{\nu}S\partial^{\mu}U\big),
\end{align}
and introduced the two auxiliary fields $U$ and $S$ through the
equations 
\begin{align}
\Box U & \equiv-R,\label{eq:BoxU}\\
\Box S & \equiv-U.\label{eq:BoxS}
\end{align}
Writing the field equations in terms of $U$ and $S$ allows us to
work with a local formulation of the theory~\cite{Maggiore:2014sia}.
In order to solve Eq. (\ref{eqmRiBox2R}) we need to first solve Eqs.
(\ref{eq:BoxU}) and (\ref{eq:BoxS}). The general solutions for $U$
and $S$ are given by 
\begin{align}
U & \equiv U_{\text{hom}}-\Box_{\text{ret}}^{-1}R,\label{eq:udefin}\\
S & \equiv S_{\text{hom}}-\Box_{\text{ret}}^{-1}U,\label{eq:uvdefin}
\end{align}
with $U_{\text{hom}}$ and $S_{\text{hom}}$ the solutions to the
homogeneous equations 
\begin{equation}
\Box U_{\text{hom}}=0,\hspace{10mm}\Box S_{\text{hom}}=0,\label{UShom}
\end{equation}
and $\Box_{\text{ret}}^{-1}$ the inverse of the retarded d'Alembertian
operator. The equivalent local form of the theory then depends on
the choice of $U_{\text{hom}}$ and $S_{\text{hom}}$. The \textit{ad hoc} choice of a retarded Green 
function in the definition of inverse d'Alembertian operator $\Box^{-1}$ will ensure causality 
(for details see e.g. Ref.~\cite{Zhang:2016ykx}). Note, however, that it has been argued that causality 
can emerge automatically if one considers only in-in (observable) vacuum expectation values \cite{Calzetta:1986ey,Jordan:1986ug,Foffa:2013vma}.

Let us now turn to our studies of the cosmology of the model. We will
assume a flat Friedmann-Lema\^{i}tre-Robertson-Walker (FLRW) metric
\begin{equation}
\text{d}s^{2}=-\text{d}t^{2}+a{}^{2}\left(t\right)\text{d}\vec{x}^{2},
\end{equation}
wtih $t$ the cosmic time and $a$ the scale factor.

Solving the field equations for this metric yields the evolution equations
(equivalent to the Friedmann equation)~\cite{Maggiore:2014sia}

\begin{eqnarray}
h^{2} & = & \dfrac{\Omega_{\text{M}}^{0}e^{-3N}+\Omega_{\text{R}}^{0}e^{-4N}+\left(\gamma/4\right)U^{2}}
{1+\gamma\left(-3V-3V'+(1/2)V'U'\right)},\label{eq:h2ithomega}\\
U'' & = & 6\left(2+\xi\right)-\left(3+\xi\right)U',\label{eq:u1}\\
V'' & = & h^{-2}U-\left(3+\xi\right)V',\label{eq:v1}
\end{eqnarray}
in terms of the auxiliary fields $U$ and $V\equiv H_{0}^{2}S$, and their derivatives, with $H_{0}$ the
present Hubble rate. Additionally, we have assumed the Universe to
be filled with matter and radiation, with present density parameters
$\Omega_{\text{M}}^{0}$ and $\Omega_{\text{R}}^{0}$, respectively,
and have defined the quantities 
\begin{equation}
\gamma\equiv\frac{m^{2}}{9H_{0}^{2}},\hspace{10mm}h\equiv\frac{H}{H_{0}},\hspace{10mm}\xi\equiv\dfrac{h'}{h}\,,\label{eq:Y}
\end{equation}
where a prime denotes a derivative with respect to the number of $e$-foldings
$N\equiv\ln a$.

The evolution of the total energy density can be parametrized in terms
of an effective equation of state \cite{2010deto.book.....A} 
\begin{align}
w_{\text{eff}} & =-1-\frac{2}{3}\frac{h'}{h}=-1-\frac{2}{3}\xi\,.\label{eq:weffdef-1}
\end{align}
The evolutions of the matter, radiation and dark energy components
contributing to $w_{\text{eff}}$ follow from the conservation of
the energy-momentum tensor, 
\begin{equation}
\Omega'_{\text{M}}+\left(3+2\xi\right)\Omega_{\text{M}}=0,\hspace{5mm}\Omega'_{\text{R}}+\left(4+2\xi\right)\Omega_{\text{R}}=0,\hspace{5mm}\Omega'_{\text{DE}}+\left(3+3w_{\text{DE}}+2\xi\right)\Omega_{\text{DE}}=0,\label{eq:conservation}
\end{equation}
with 
\begin{equation}
\xi=\frac{-4\Omega_{\text{R}}-3\Omega_{\text{M}}+3\gamma\left(h^{-2}U+U'V'-4V'\right)}{2\left(1-3\gamma V\right)}.\label{eq:dittadef}
\end{equation}
Combining the conservation equations~(\ref{eq:conservation}) and
taking into account the cosmic sum rule, 
\begin{equation}
\Omega_{\text{DE}}=1-h^{-2}\left(\Omega_{\text{M}}^{0}e^{-3N}+\Omega_{\text{R}}^{0}e^{-4N}\right)=\gamma\left(\dfrac{1}{4}h^{-2}U^{2}+3V+3V'-\dfrac{1}{2}V'U'\right),
\end{equation}
we obtain the dark energy equation of state, 
\begin{equation}
w_{\text{DE}}=\frac{\gamma\left(4(U+3)-(U+2)U'\right)V'+U(\gamma V(\Omega_{\text{R}}+3)-\Omega_{\text{DE}})+4(3\gamma V-\Omega_{\text{DE}})}{U(1-3\gamma V)\Omega_{\text{DE}}}.
\end{equation}

\section{Phase space and dynamical analysis}

\label{sec:dynanalysis}

In order to perform the dynamical analysis of the model, it is convenient
to rewrite the second order differential equations \eqref{eq:u1}
and \eqref{eq:v1} in a first order form. To do this, we introduce
two new fields $Y_{1}$ and $Y_{2}$ defined as $Y_{1}\equiv U'$
and $Y_{2}\equiv V'$. We can now rewrite the system as a set of six
autonomous first order differential equations 
\begin{eqnarray}
U' & = & Y_{1},\label{eq:1set}\\
V' & = & Y_{2},\\
Y'_{1} & = & -\frac{3((U+2)Y_{1}-4(U+3))(2(3\gamma V-1)-\gamma(Y_{1}-6)Y_{2})+3(U+4)(Y_{1}-6)\Omega_{\text{M}}+4(U+3)(Y_{1}-6)\Omega_{\text{R}}}{2U(3\gamma V-1)},\label{eq:lastset1}\\
Y'_{2} & = & -\frac{Y_{2}(6(3\gamma V-1)-3\gamma(Y_{1}-4)Y_{2}+3\Omega_{\text{M}}+4\Omega_{\text{R}})}{2(3\gamma V-1)}\label{eq:lastset2}\\
 & - & \dfrac{(2(3\gamma V-1)+3\gamma Y_{2})(2(3\gamma V-1)-\gamma(Y_{1}-6)Y_{2}+2(\Omega_{\text{M}}+\Omega_{\text{R}}))}{\gamma U(3\gamma V-1)},\nonumber \\
\Omega'_{\text{M}} & = & -\frac{\Omega_{\text{M}}(U(3(3\gamma V-1)-3\gamma(Y_{1}-4)Y_{2}+3\Omega_{\text{M}}+4\Omega_{\text{R}})+12(3\gamma V-1)+12\left(\Omega_{\text{M}}+\Omega_{\text{R}}\right)-6\gamma(Y_{1}-6)Y_{2})}{U(3\gamma V-1)},\label{eq:lastset3}\\
\Omega'_{\text{R}} & = & -\frac{\Omega_{\text{R}}(U(4\left(3\gamma V-1\right)-3\gamma(Y_{1}-4)Y_{2}+3\Omega_{\text{M}}+4\Omega_{\text{R}})+12(3\gamma V-1)+12\left(\Omega_{\text{M}}+\Omega_{\text{R}}\right)-6\gamma(Y_{1}-6)Y_{2})}{U(3\gamma V-1)}.\label{eq:lastset4}
\end{eqnarray}
A quick look at Eqs.~\eqref{eq:1set}-\eqref{eq:lastset4} reveals
that they are not invariant under $U\rightarrow U+U_{\text{hom}}$
and $\rho\rightarrow\rho+\Lambda$, where $\rho$ is the energy density
of the system. Contrary to the nonlocal models considered in Ref.~\cite{Foffa:2013vma},
non-zero and constant values of $U_{\text{hom}}$ \textit{are not}
equivalent to a cosmological constant. The main purpose of this work
is a complete characterization of the system \eqref{eq:1set}-\eqref{eq:lastset4}
for arbitrary values of $U_{\text{hom}}$ and $V_{\text{hom}}$. As
argued in Ref.~\cite{Maggiore:2014sia}, each choice of $U_{\text{hom}}$
and $V_{\text{hom}}$ in Eq.~\eqref{UShom} (note that $S$ and $V$
are the same up to a constant factor) corresponds to the choice of
one and only one boundary condition in the nonlocal formulation of
the theory. Different initial conditions, and therefore different
solutions, should be associated with different nonlocal models. The
qualitative analysis of Eqs.~\eqref{eq:h2ithomega}-\eqref{eq:v1}
will allow us to understand which of these models are phenomenologically
viable.

\subsection{Critical points and evolution paths: Numerical analysis}

\label{sec:PS_dyn}

The fixed points of the dynamical system~\eqref{eq:1set}-\eqref{eq:lastset4}
are those at which all the first derivatives on the left-hand side
of the equations vanish. In some cases though, one can have fixed
surfaces instead of fixed points, that
is, only a subset of variables is constant. In order to go from
the fixed surfaces to fixed points (in
a lower dimensional phase space) one has to perform an appropriate
variable transformation (cf. Appendix \ref{appendix1} for details
regarding the treatment of fixed lines). By following this procedure,
we obtain the five nontrivial fixed points/surfaces
I-V listed in Table~\ref{tab:fixedpoints}. The values of the dynamical
variables of the system (i.e. the quantities $U$, $V$, $U'$, $V'$,
$\Omega_{\text{M}}$, and $\Omega_{\text{R}}$) are given for each
point, as well as the value of the effective equation of state parameter
$w_{\text{eff}}$. As reflected in the table, we find two attractors
and three saddle points.%
\footnote{One should note that for point III, Eqs. \eqref{eq:1set}-\eqref{eq:lastset4}
may seem to be singular in the limit $V\rightarrow1/3\gamma$. This
is, however, not the case, as in the limit $V'=\Omega_{\text{M}}=\Omega_{\text{R}}=0$
and $U'=4$, the divergent factor is canceled out.%
} 
\begin{table}[t]
\centering{}%
\begin{tabular}{|c|c|c|c|c|c|c|c|c|}
\hline 
Point  & $U$  & $V$  & $U'$  & $V'$  & $\Omega_{\text{M}}$  & $\Omega_{\text{R}}$  & $w_{\text{eff}}$  & Type \tabularnewline
\hline 
\hline 
I  & $\widetilde{U}$  & $(1-\widetilde{\Omega}_{\text{R}})/(3\gamma)$  & $0$  & $0$  & $0$  & $\widetilde{\Omega}_{\text{R}}$  & $1/3$  & Saddle \tabularnewline
\hline 
II  & $2N+\widetilde{U}$  & $(1-\widetilde{\Omega}_{\text{M}})/(3\gamma)$  & $2$  & $0$  & $\widetilde{\Omega}_{\text{M}}$  & $0$  & $0$  & Saddle \tabularnewline
\hline 
III  &$+\infty$  & $1/(3\gamma)$  & $4$  & $0$  & $0$  & $0$  & $-1$  & Attractor \tabularnewline
\hline 
IV  & $4N+\widetilde{U}$  & $\pm\infty$  & $4$  & $\pm\infty$  & $0$  & $0$  & $-1$  & Saddle  \tabularnewline
\hline 
V  & $-3$  & $\pm\infty$  & $0$  & $4V$  & $\mp\infty$  & $0$  & $1/3$  & Attractor\tabularnewline
\hline 
\end{tabular}\protect\caption{\label{tab:fixedpoints}Critical points of the dynamical system \eqref{eq:1set}-\eqref{eq:lastset4}.
The quantities $\widetilde{\Omega}_{\text{M}}$, $\widetilde{\Omega}_{\text{R}}$,
and $\widetilde{U}$ stand, respectively, for some constant values
of $\Omega_{\text{M}}$, $\Omega_{\text{R}}$ and $U$.}
\end{table}

The behavior of the solutions around each of the critical points can
be determined by using the standard phase-space analysis methods.
Although the set of equations \eqref{eq:1set}-\eqref{eq:lastset4}
is nonlinear, the system behaves linearly in the vicinity of each
critical point, provided that the point is isolated and the Jacobian
at the point is invertible.%
\footnote{This argumentation holds only if the fixed point of the linearized
system is not a center-type point.%
} The linearization of Eqs. \eqref{eq:1set}-\eqref{eq:lastset4} in
the vicinity of each fixed point gives rise to a set of linear equations,
which can be generically written in a matrix form $\mathbf{X'}={\bf A}\cdot\mathbf{X}$,
with ${\bf A}$ a $6\times6$ matrix and $\mathbf{X}=\{U,V,U',V',\Omega_{\text{M}},\Omega_{\text{R}}\}$.
The behavior of the system around each critical point is determined
by the eigenvalues of the corresponding ${\bf A}$ matrix. The results
are summarized in the last column of Table \ref{tab:fixedpoints}
(cf. Appendix \ref{appendix1} for details).

\begin{figure}
\centering{} \includegraphics[scale=0.8]{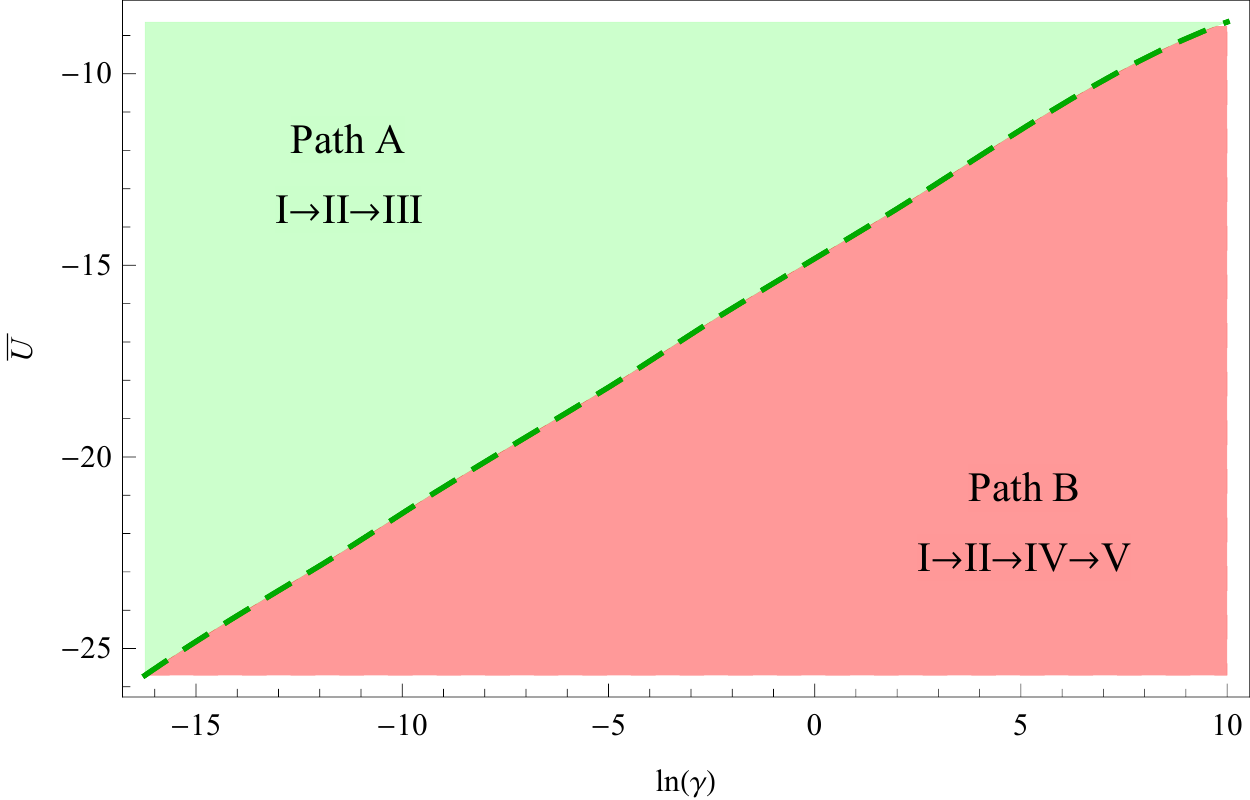} \protect\protect\protect\protect\caption{\label{fig:ucritgamma} The two evolution paths A and B for the background
cosmology of $R\,\Box^{-2}R$ gravity, in terms of the initial value
$U_{0}$ of the auxiliary field $U$ and the value of $\gamma\equiv\frac{m^{2}}{9H_{0}^{2}}$.
The diagonal line depicts the critical value $\bar{U}$ as a function
of $\gamma$. The green (red) region corresponds to the realizations
of path A (B).}

\label{ubpath} 
\end{figure}

The precise interpolation of the solutions between the critical points
I-V depends on the initial conditions, and in particular, on the relation
between the initial value $U_{0}$ and a $\gamma$-dependent critical
value $\bar{U}$ that we obtain numerically for the case $\Omega_{\text{M}}^{0}=0.3$,\footnote{We will recover 
analytically the $\gamma$-dependent part of this equation in Sec.~\ref{sec:evolpaths}.}
\begin{equation}
\bar{U}(\gamma)\simeq-14.82+0.67\log\gamma\,,\label{eq:ugamma}
\end{equation}
and that is valid in the range illustrated in Fig. \ref{fig:ucritgamma}.
We can distinguish two kinds of trajectory. If $U_{0}$ (the initial
value of $U$) is bigger than $\bar{U}$, the system follows the sequence
I$\to$II$\to$III. In the opposite case, it follows the I$\to$II$\to$IV$\to$V
sequence. We will refer to these two possibilities as path A and path
B, respectively (see Fig. \ref{ubpath}). The previous work on this
model, i.e. Ref.~\cite{Dirian:2014bma}, has focused on the particular
case of path A, as we discuss in detail below.

\subsubsection{Path A}

\label{sec:pathA}

\begin{figure}
\begin{centering}
\hspace{1cm}\includegraphics[height=5cm]{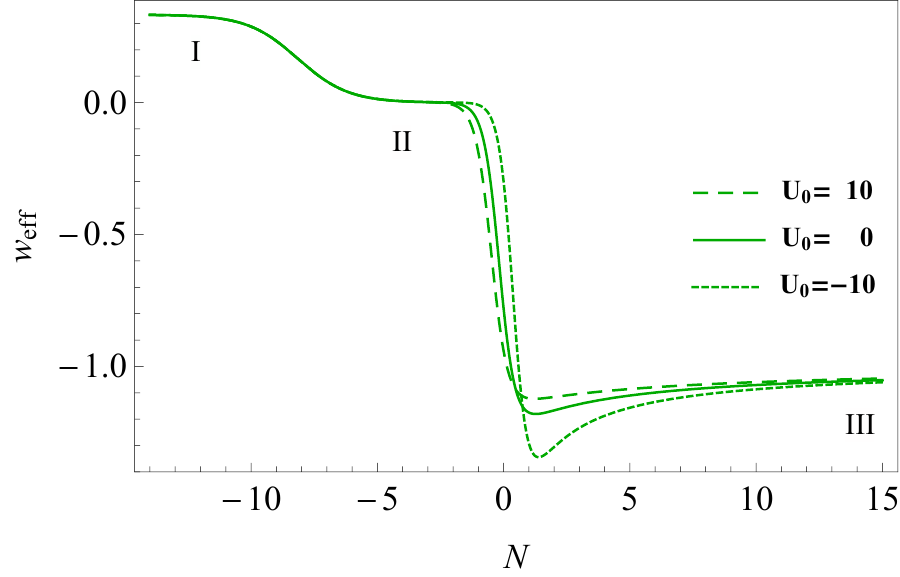} \hspace{0.8cm}\includegraphics[height=5cm]{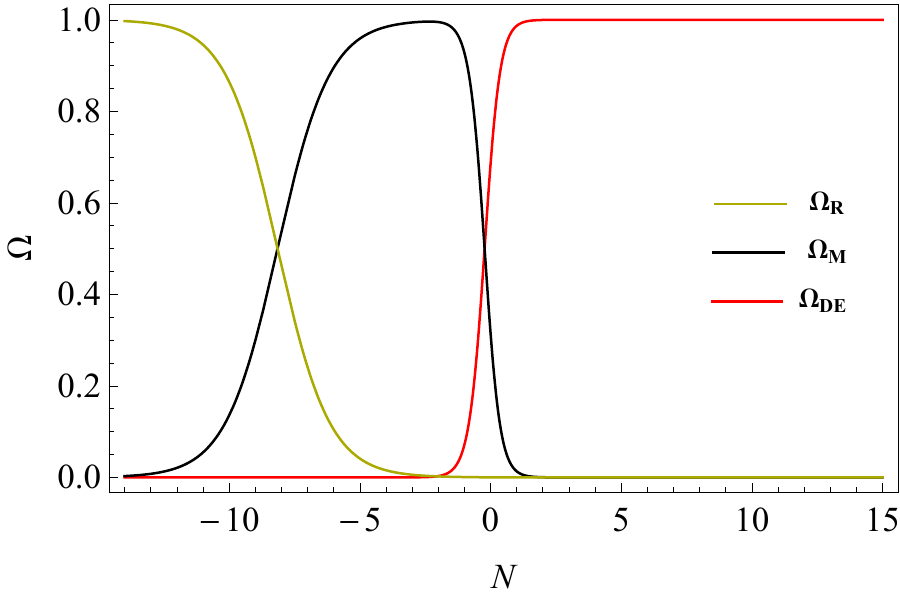} 
\par\end{centering}

\protect\protect\protect\protect\protect\caption{\label{fig:Vandweff} (Left) Evolution of the effective equation of
state $w_{\text{eff}}$ as a function of $N\equiv\ln a$ for path
A. (Right) Evolution of the density parameters $\Omega_{\text{M}},\Omega_{\text{R}}$,
and $\Omega_{\text{DE}}$ for the same path with $U_{0}=0$. In both
plots we have fixed $V_{0}=0$.}
\end{figure}

\begin{figure}
\begin{centering}
\includegraphics[height=5cm]{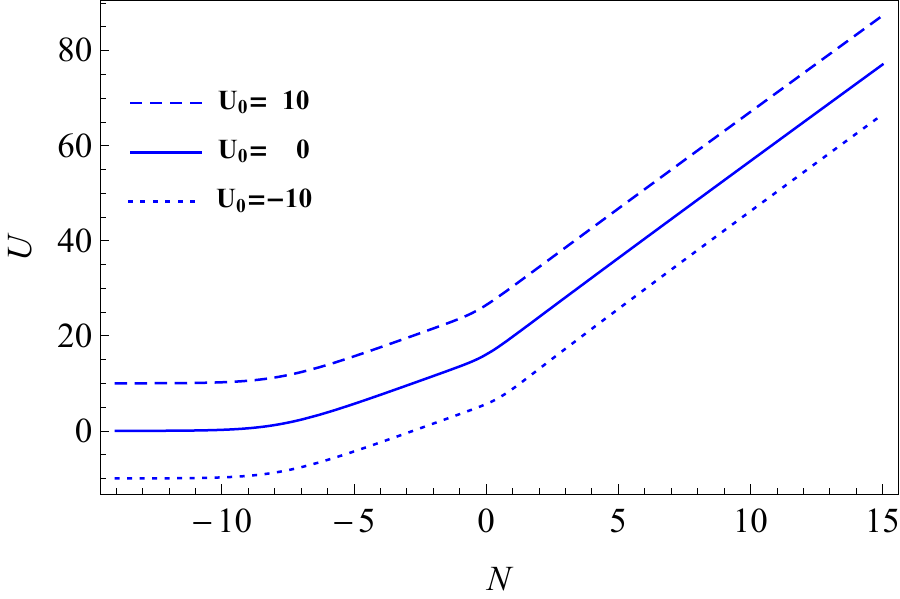}\hspace{0.8cm}
\includegraphics[height=5cm]{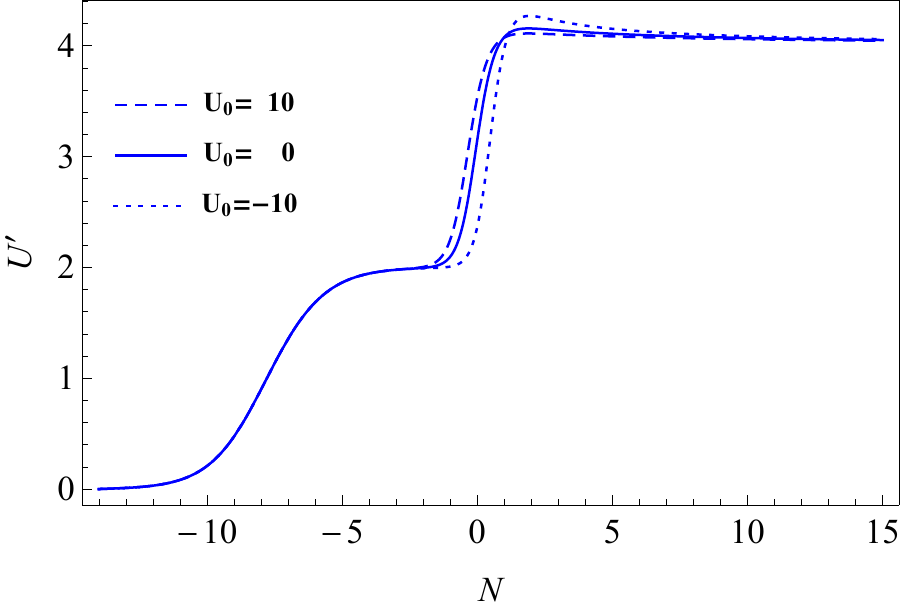} \\
 \vspace{0.3cm}
 \includegraphics[height=5cm]{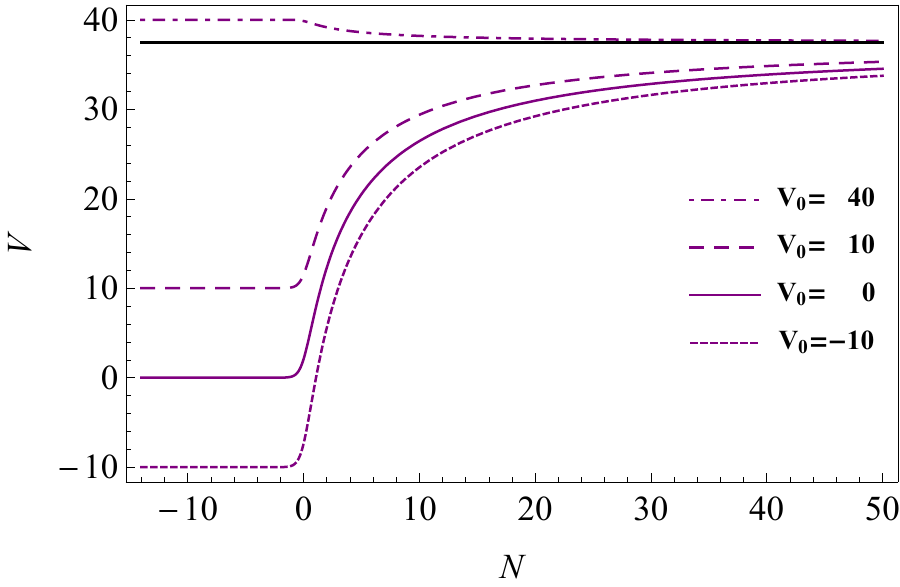} \hspace{0.8cm}
 \includegraphics[height=5cm]{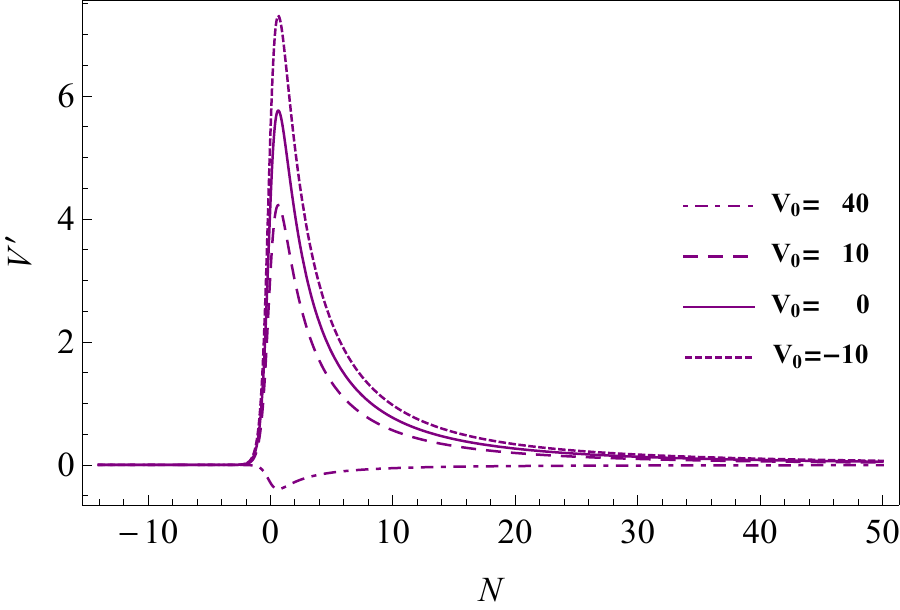}\end{centering}
\protect\protect\protect\protect\protect\caption{\label{fig:Firstpointuu} Evolution of the auxiliary fields $U$ and
$V$, and their derivatives with respect to $N\equiv\ln a$, $U'$
and $V'$, for path A. In the plots of $U$ and $U'$ we have fixed
$V_{0}=0$, and in the plots of $V$ and $V'$ we have fixed $U_{0}=0$.}
\end{figure}

The numerical behavior of the dynamical system along path A is shown
in Figs.~\ref{fig:Vandweff} and \ref{fig:Firstpointuu}. Note that
in Fig.~\ref{fig:Vandweff} we have fixed $V_{0}=0$. This choice
can be made without loss of generality due to the attractor behavior
of point III.

As can be clearly seen in Fig.~\ref{fig:Vandweff}, the saddle points
I and II correspond to intermediate radiation- and matter-dominated
eras. The transition to the attractor point III proceeds through a
transient phantom regime with $w_{\text{eff}}<-1$. This kind of behavior
was first recognized in Ref.~\cite{Dirian:2014bma} where the authors
considered the solution of the dynamical system \eqref{eq:h2ithomega}-\eqref{eq:v1}
for a specific choice of the initial conditions ($U_{0}=0$, $V_{0}=0$)
and derived a lower bound for the effective equation of state ($-1.14\leq w_{\text{eff}}<-1$).
As shown in Fig.~\ref{fig:Vandweff}, this bound is not robust under
variations of the initial conditions. General choices of $U_{0}$
can lead to a stronger phantom regime (or even to its complete disappearance,
cf. Sec. \ref{pathB}). Note also that the particular choice of initial
conditions in the MM model rests on the assumption of a vanishing
Ricci scalar prior to matter-radiation equality, or in others words,
on the existence of a perfect radiation-dominated era. However, the
accuracy and redshifts for which this assumption holds depend on the
thermal history of the Universe. As shown in Appendix \ref{appendix2},
if the initial MM conditions were set for instance at the end of inflation/reheating, one should
expect nonvanishing values of $U_{0}$ at the number of $e$-foldings at 
which the MM initial conditions are usually implemented
($N\simeq -14$)~\cite{Dirian:2014ara}.

In spite of the asymptotic approach of the effective equation of state
to $w_{\text{eff}}=-1$, the attractor point III \textit{should not}
be identified, \textit{sensu stricto}, with a de Sitter point. For
a solution to be de Sitter, the Hubble parameter around this solution
should remain constant \textcolor{black}{(or, more generally, the
Ricci scalar $R$ should be constant).} This is certainly not the
case here. Indeed, the Hubble rate $H(N)$ becomes infinitely large
when $N\rightarrow\infty$. This unusual behavior can be easily understood
by considering the consistency of Eqs.~(\ref{eq:u1}), (\ref{eq:v1}),
and (\ref{eq:weffdef-1}) at the fixed point III.%
\footnote{In order for Eq.~\eqref{eq:v1} to be consistent at III, we must
require $H\rightarrow\infty$ faster than $U$.%
}

\subsubsection{Path B}

\label{sec:pathB}

\label{pathB}

\begin{figure}
\begin{centering}
\includegraphics[height=5cm]{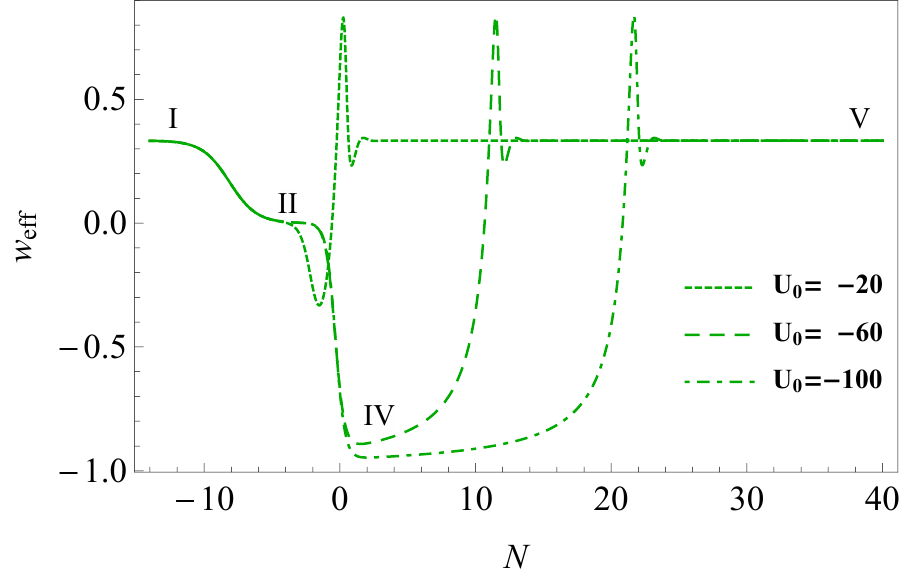}\hspace{0.8cm}
\includegraphics[height=5cm]{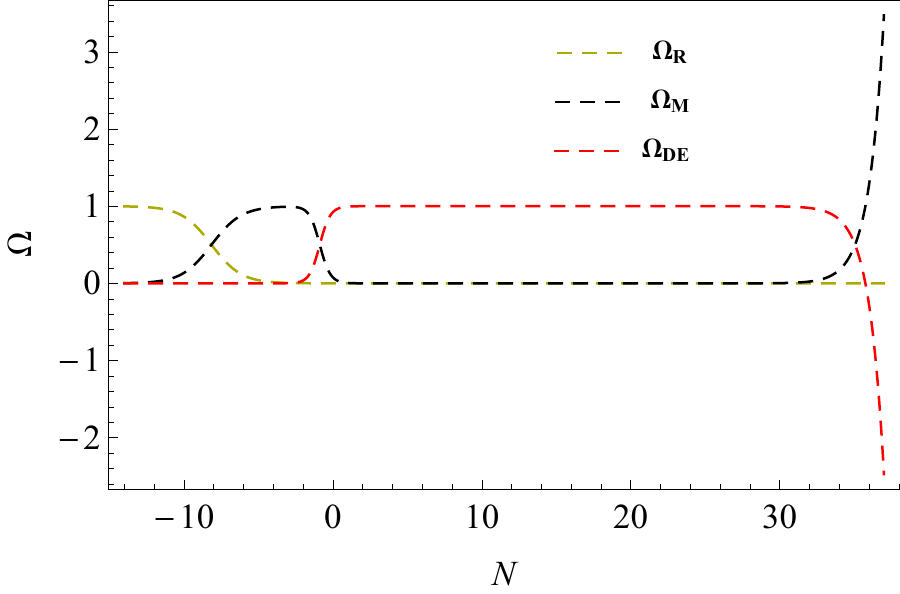}
\end{centering}
\protect\protect\protect\protect\protect\caption{\label{fig:weffandommega3} (Left) Evolution of the effective equation
of state $w_{\text{eff}}$ as a function of $N\equiv\ln a$ for path
B. (Right) Evolution of the density parameters $\Omega_{\text{M}}$,
$\Omega_{\text{R}}$, and $\Omega_{\text{DE}}$ for the same path
with $U_{0}=-60$. In both plots we have fixed $V_{0}=0$.}
\end{figure}

\begin{figure}
\begin{centering}
\includegraphics[height=5cm]{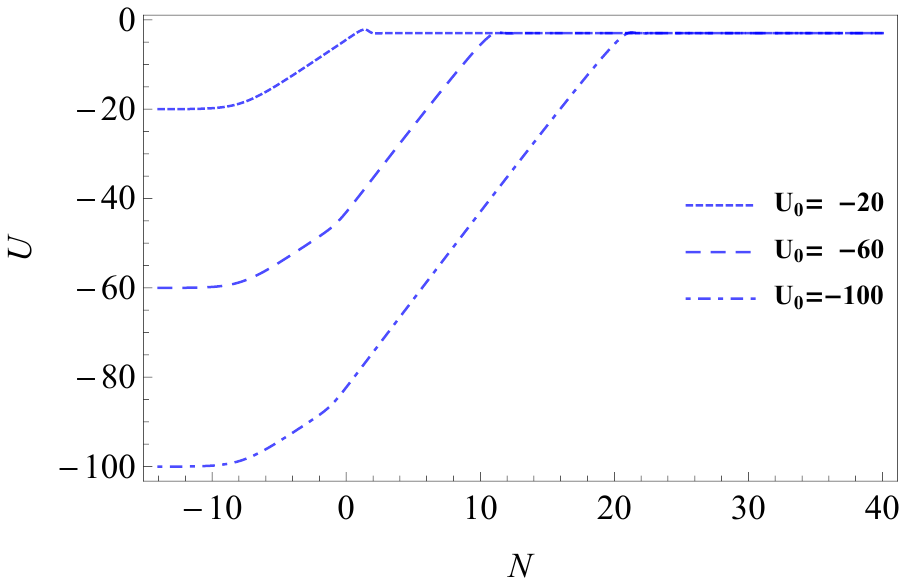}\hspace{0.8cm}
\includegraphics[height=5cm]{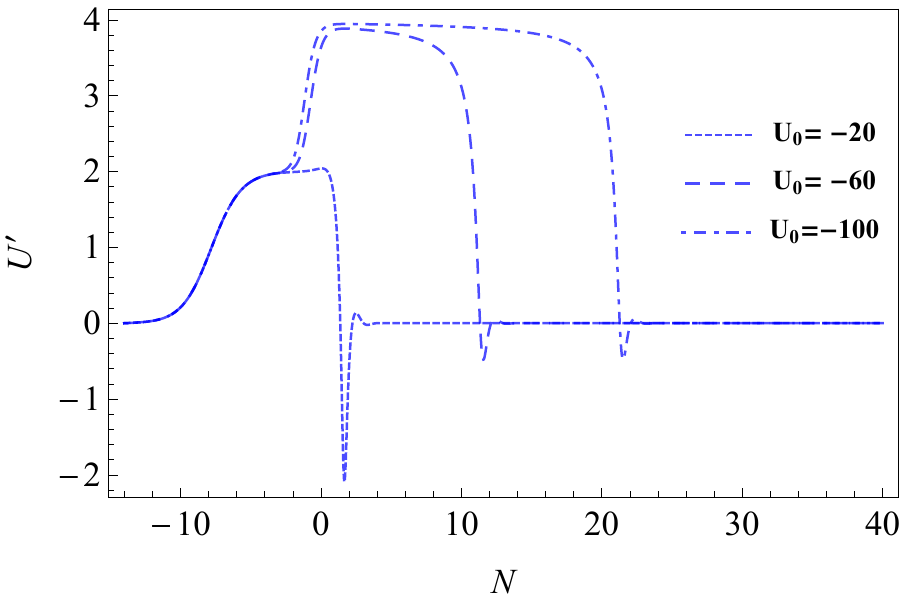}\\
\vspace{0.3cm}
\includegraphics[height=4.9cm]{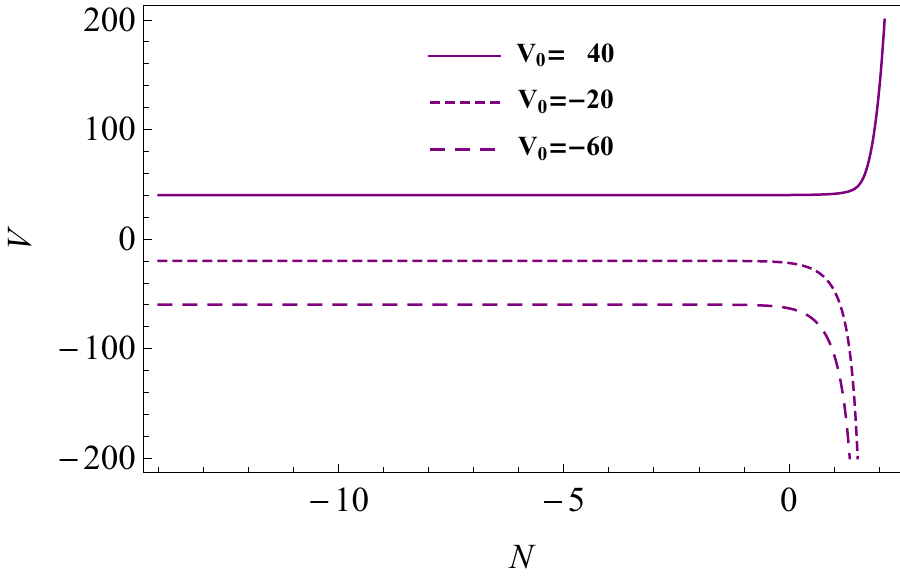}\hspace{0.8cm}
\includegraphics[height=4.9cm]{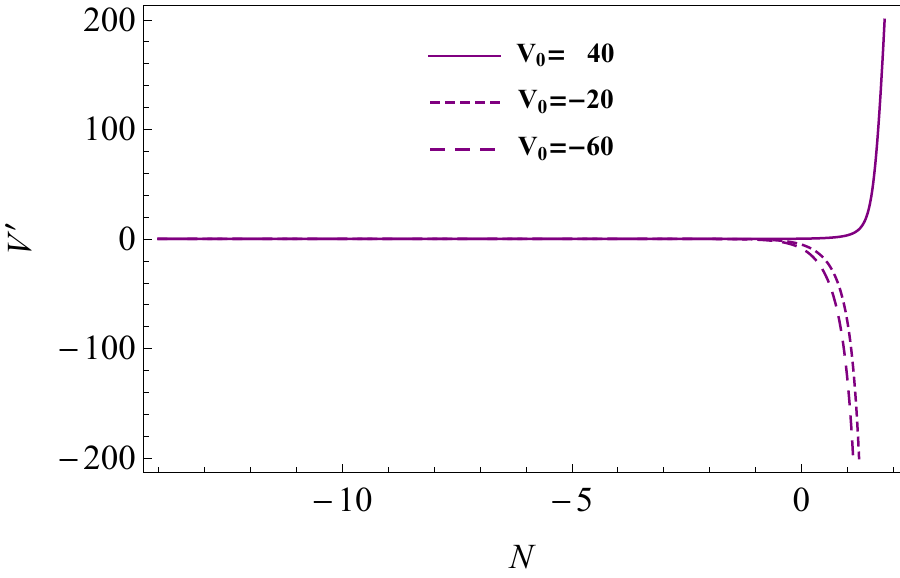}
\end{centering}
\protect\protect\protect\protect\protect\caption{\label{fig:UVfin3}Evolution of the auxiliary fields $U$ and $V$,
and their derivatives with respect to $N\equiv\ln a$, $U'$ and $V'$,
for path B. In the plots of $U$ and $U'$ we have fixed $V_{0}=0$,
and in the plots of $V$ and $V'$ we have fixed $U_{0}=-20$.}
\end{figure}

The numerical evolution of the dynamical system along path B is shown
in Figs.~\ref{fig:weffandommega3} and \ref{fig:UVfin3}. Note that
in Fig.~\ref{fig:weffandommega3} we have fixed $V_{0}=0$. This
can be done without loss of generality, provided that $V_{0}<1/(3\gamma)-V_{0}'$
(see the discussion below).

The initial behavior of the system coincides with that in path A.
In particular, the Universe undergoes radiation- and matter-dominated
eras while passing through the saddle points I and II. The differences
appear only when the system approaches the fixed point IV. As shown
on the left-hand side of Fig.~\ref{fig:weffandommega3}, this point
gives rise to a \textit{true} de Sitter epoch with $w_{\text{eff}}\simeq-1$
and $H(N)\simeq\textrm{constant}$. Note, however, that this point
is not an attractor but rather a saddle point. This means that the
solution stays close to the point for some period of time but eventually
moves to the final attractor, the fixed point V. In particular, the
late-time evolution depends on the value of $V_{0}$ and $V_{0}'$.
As discussed in Appendix \ref{appendix1}, if $V_{0}>1/(3\gamma)-V_{0}'$
then the system approaches the fixed point V with $\Omega_{\text{M}}\rightarrow-\infty$.
Since $\Omega_{\text{M}}$ takes negative values when $V_{0}>1/(3\gamma)-V_{0}'$,
this set of initial conditions should be discarded on general physical
grounds. On the contrary, if $V_{0}<1/(3\gamma)-V_{0}'$ we can obtain
a physically viable scenario. As shown on the right-hand side of Fig.~\ref{fig:weffandommega3},
the matter density parameter in this case is driven to $+\infty$
while the dark energy one goes to $-\infty$. This limit is acceptable
since $\Omega_{\text{DE}}$ does not represent a proper matter content
but rather an effective description of the gravitational degrees of
freedom. Note that the effective equation of state at point V approaches
the radiation-domination value $w_{\text{eff}}=1/3$, even though
there is no radiation left.%
\footnote{In fact, for all the points III, IV, and V, $\Omega_{\text{R}}\to0$.%
}

The cosmological evolution along path B requires only that $U_{0}<\bar{U}$.
The value of $U_{0}$ is in principle unbounded from below. Could
it be possible to obtain a phantom regime similar to that occurring
for path A by choosing $U_{0}\ll\bar{U}$? The answer to this question
turns out to be negative. As shown in Fig.~\ref{fig:UVfin3}, when
we increase the absolute value of $U_{0}$, the variable $U'$ approaches
a maximal value $U'_{\text{max}}=4$, stays there for some time interval
$\Delta N_{\text{max}}^{U'}(U_{0})$, and eventually falls into its
future attractor regime $U'=0$. The maximum value of $U'$ ($U'_{\text{max}}=4$)
translates, through Eq.~(\ref{eq:u1}), into a value $\xi_{\text{max}}=0$,
and as a result, $w_{\text{eff}}=-1-\frac{2}{3}\xi$ cannot be smaller
than $-1$. In other words, path B is never phantom.

\subsection{Evolution paths: Analytical results}\label{sec:evolpaths}

The novel ingredient of the local formulation of the $R\,\Box^{-2}R$
model with respect to general relativity is the presence of two ``integral
fields'' $U$ and $V$ arising from the nonlocal structure of the
theory, cf.~Eqs~(\ref{eq:udefin}) and (\ref{eq:uvdefin}). In this
subsection, we take an in-depth look at the evolution of the $2+2$
homogeneous and $1+1$ inhomogeneous modes and analytically confirm
the results obtained in Secs. \ref{sec:pathA} and \ref{sec:pathB}.

The basic building blocks of cosmological model construction are solutions
with constant effective equation of state $w_{\text{eff}}$. Assuming
$w_{\text{eff}}=w_{c}$ with $w_{c}$ a constant, and using Eqs.~\eqref{eq:udefin}
and \eqref{eq:uvdefin}, we obtain the following equations for the
$U$ and $V$ fields 
\begin{equation}
U(N)=u_{0}+u_{1}e^{-\frac{3}{2}(1-w_{c})N}+\frac{2(1-3w_{c})}{1-w_{c}}N,\label{eq:solu}
\end{equation}
\begin{equation}
V(N)=\frac{2e^{3(1+w_{c})N}}{9(1+w_{c})(3+w_{c})}\left(u_{0}-\frac{2(1-3w_{c})(5+3w_{c})}{3(1-w_{c}^{2})(3+w_{c})}-\frac{2(1-3w_{c})}{1-w_{c}}N\right)+\frac{2e^{\frac{3}{2}(1+3w_{c})N}}{9(1+w_{c})(1+3w_{c})}u_{1}+v_{0}+v_{1}e^{-\frac{3}{2}(1-w_{c})N},\label{eq:solv}
\end{equation}
with $u_{0}$, $u_{1}$, $v_{0}$, and $v_{1}$%
\footnote{The values $u_{0}$ and $v_{0}$, that are set at $N=0$ according to the solutions (\ref{eq:solu}) and (\ref{eq:solv}), should not in general be confused with
$U_{0}$ and $V_{0}$, the initial values for $U$ and $V$ set at an early radiation-dominated epoch.%
} integration constants set at $N=0$. These equations reveal
that the inhomogeneous modes disappear if and only if $w_{c}=1/3$.
As a default, $w_{c}=1/3$ is the only constant equation of state
giving rise to an attractor solution.%
\footnote{Eqs.~\eqref{eq:solu} and \eqref{eq:solv} are exact and model-independent
solutions as long as we can assume that $w_{\text{eff}}$ is a constant.%
} Note also that for $-3<w_{c}<1$ the fastest-growing exponent in
Eq.~(\ref{eq:solv}) appears in the first term, which is controlled
by $u_{0}$ only. Taking this into account, we will mostly focus on
variations of $u_{0}$ in what follows.

Let us first consider a solution within radiation domination, like
that taking place around the fixed point I. The growing modes in this
case are given by $U\rightarrow u_{0}$ and $V\rightarrow\frac{1}{20}e^{4N}u_{0}$.
If we start with an initial condition $\Omega_{\text{R}}=1$, the
numerator of Eq.~\eqref{eq:h2ithomega} tells us that the nonlocal
corrections take over at $N_{{\rm NL}}=-\frac{1}{4}\log{(\frac{\gamma u_{0}}{4})}$
$e$-foldings. Thus, the radiation-dominated Universe in the MM model
is stable if and only if we have exactly the minimal boundary
condition prescription.

In a realistic cosmology we should also consider a matter-dominated
epoch following the radiation-domination era. This matter-dominated
era happens around the critical point II. For $w_{c}=0$, the inhomogeneous modes in Eqs.~\eqref{eq:solu}
and \eqref{eq:solv} survive and the solution is necessarily unstable.
The number of $e$-foldings $N_{{\rm NL}}$ at which the nonlocal
corrections take over is again dictated by the numerator of Eq.~(\ref{eq:h2ithomega}).
At $N_{{\rm NL}}$ $e$-foldings, the sign of the fastest-growing
mode is positive if %
\footnote{This formula is approximate because when $\Omega_{\text{M}}=1/2$,
$w_{c}=0$ is not exact.%
} 
\begin{equation}\label{u0crit}
 u_{0}>-\left(\frac{10}{9}+\frac{4}{3}\log{\frac{9}{5}}\right)+\frac{2}{3}\log{\gamma}.
\end{equation}
 As we will confirm below, one should expect this sign to determine the evolution of the
system beyond point II. Note that Eq.~\eqref{u0crit} can be 
translated into a bound on the value of $U(N_*)$ at any given number of $e$-foldings $N_*$ 
by noticing that 
\begin{equation}
 u_{0}=U(N_*)-2\frac{{(1-3w_{c})N_*}}{1-w_{c}}-\frac{4(1-3w_{c})}{3(1-w_{c})^2}.
\end{equation}
In particular, for matter-radiation equality ($N_*=-8.1$), we get $U(N_*)>-15.65+\frac{2}{3}\log{\gamma}$. Note that this is close to 
the numerical value of  $\bar U$ found in Eq.~\eqref{eq:ugamma}.

The above two cases constitute the only possibilities for realizing
a constant equation of state $w_{\text{eff}}=w_{c}$ in a universe
with nonvanishing and minimally coupled radiation and dust components.
In what follows, we will consider vacuum solutions with $\Omega_{\text{R}}=\Omega_{\text{M}}=0$.
Combining Eqs.~\eqref{eq:h2ithomega} and \eqref{eq:v1} we get 
\begin{equation}
UV''-2U'V'=-\frac{3}{2}\left(U-w_{c}U+8\right)V'-12V+\frac{4}{\gamma}\,.\label{consistency}
\end{equation}
As in the nonvacuum case, the attractor solutions can be associated
only to an effective equation of state $w_{c}=1/3$. The fixed point
V falls into this category. Indeed, when we set $u_{1}=v_{1}=0$,
Eq. (\ref{consistency}) reduces to 
\begin{equation}
e^{4N}\left(u_{0}+3\right)-\frac{\gamma}{4u_{0}}\left(1-3\gamma v_{0}\right)=0,
\end{equation}
which allows for the solutions 
\begin{equation}
U(N)=-3,\hspace{10mm}V(N)=\frac{1}{3\gamma}-\frac{3}{20}e^{4N}.\label{sol5}
\end{equation}

Equations~\eqref{sol5} are exact solutions of the system of Eqs.
(\ref{eq:solu}), (\ref{eq:solv}), and (\ref{consistency}). Note, however,
that there can still be approximate solutions. Consider temporary
regions with $w_{c}\approx-1$, as those appearing around points III
and IV. In these regions, Eqs. (\ref{eq:solu}) and (\ref{consistency})
give $U'(N)\rightarrow4$ and $V(N)=1/(3\gamma)$, but one should
be aware that Eq. (\ref{eq:solv}) is only valid for a constant $w_{\rm eff}$.
If $U(N_{\text{NL}})<-3$ when this solution is reached at $N_{\text{NL}}$,
the trajectory will hit the aforementioned attractor with $w_{c}=1/3$
and stay there; this is then part of what we called
path B, cf. Sec. \ref{sec:pathB}. If $U(N_{\text{NL}})>-3$, the
evolution will continue in the phase with $w_{c}\approx-1$ and $U'=4$;
this phase belongs to path A, cf. Sec. \ref{sec:pathA}.

To summarize, the post-matter-dominated Universe reaches an accelerating
stage with $w_{c}\approx-1$ which goes on until $U(N)=u_{0}+4N=-3$.
If $u_{0}>-3$, this period extends forever. Otherwise, one can prolong
the transient acceleration for $\Delta N$ $e$-foldings by lowering
the initial value of $U_0\rightarrow U_0-4\,\Delta N$.

The results of this subsection are in agreement with what we studied
in greater detail in the previous subsections, namely the stability
analysis in the phase-space formulation and the numerical integration
of the field equations.

\section{Constraints from supernovae data}

\label{sec:SNe}

\begin{figure}
\begin{centering}
\includegraphics[height=5cm]{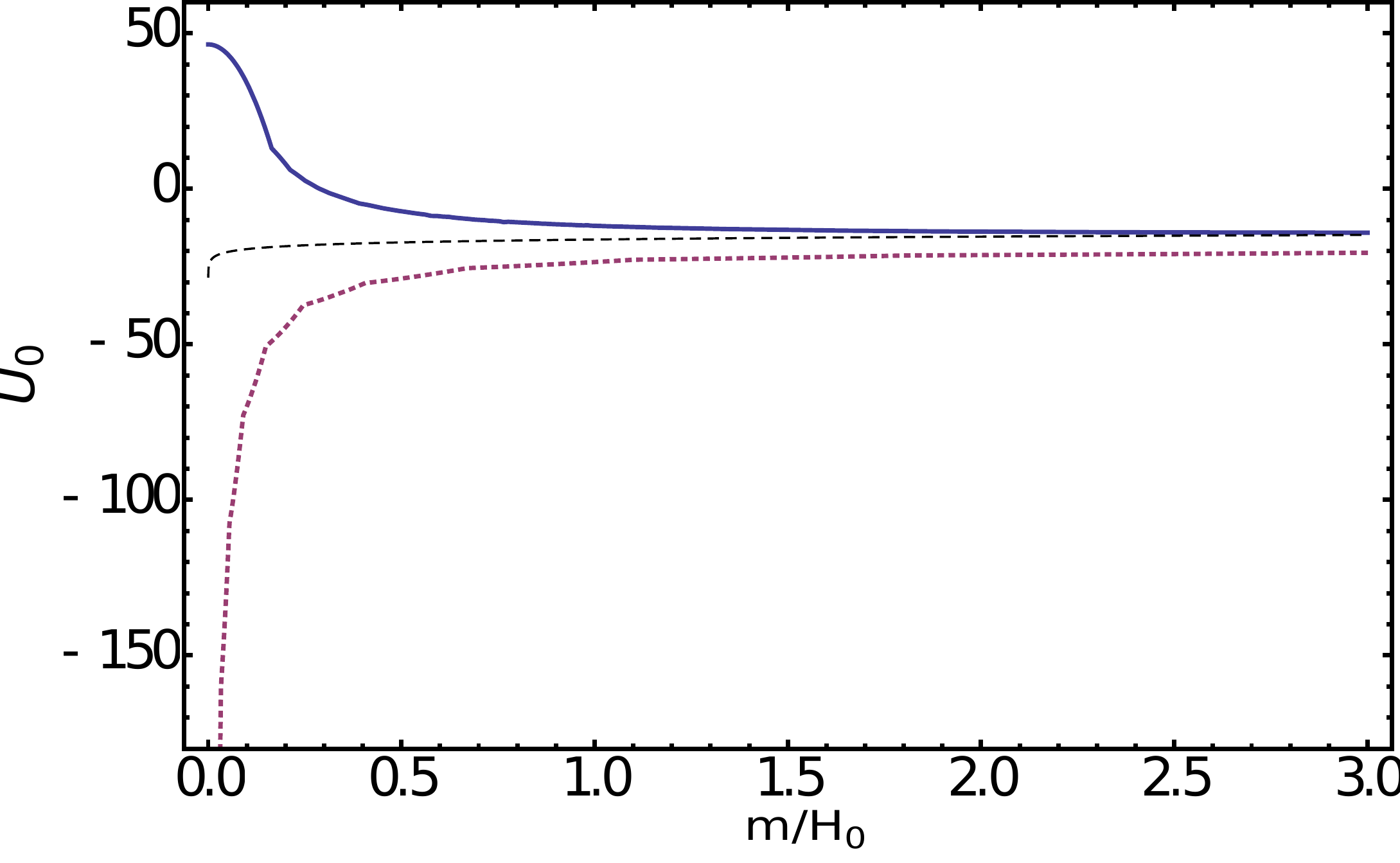} 
\par\end{centering}

\centering{}\protect\protect\protect\protect\protect\caption{\label{fig:Value-of-}Value of $U_{0}$ needed to reach path A (blue
curve) or path B (red dotted curve) when $\Omega_{\text{M}}^{0}=0.3$.
For large $m$, the two curves converge to $\bar{U}$, represented
by the intermediate dashed curve, which follows Eq. (\ref{eq:ugamma}).}
\end{figure}

Since both paths A and B realize cosmologies that are in principle
viable (i.e., they contain a sequence of proper radiation-, matter-,
and dark-energy-dominated eras), we need to compare both to observations.
Here, we assume as free parameters, $m$ in units of $H_{0}$ and
the present matter density parameter $\Omega_{\text{M}}^{0}$, and
fix $\Omega_{\text{R}}^{0}=4.15\cdot10^{-5}h^{-2}$ and $V_{0}=0$.
The initial condition deep in the radiation era, that we choose arbitrarily
as $U_{0}\equiv U(N=-14)$, is fixed by the requirement that we reach
$\Omega_{\text{M}}^{0}$ today. In practice, for every point $\lbrace{m,\Omega_{\text{M}}^{0}\rbrace}$
in the parameter space, we vary iteratively $U_{0}$ until we find
$\Omega_{\text{M}}^{0}$ at $N=0$. Since there are two possible paths,
we find two values of $U_{0}$ for every choice of parameters. 
The
particular choice of $U_{0}$ as a function of $m$ when $\Omega_{\text{M}}^{0}=0.3$
is presented in Fig. \ref{fig:Value-of-}. For large $m$, paths
A and B lead to a common behavior, and their initial condition $U_{0}$
also converges. 

\begin{figure}
\begin{centering}
\includegraphics[height=6cm]{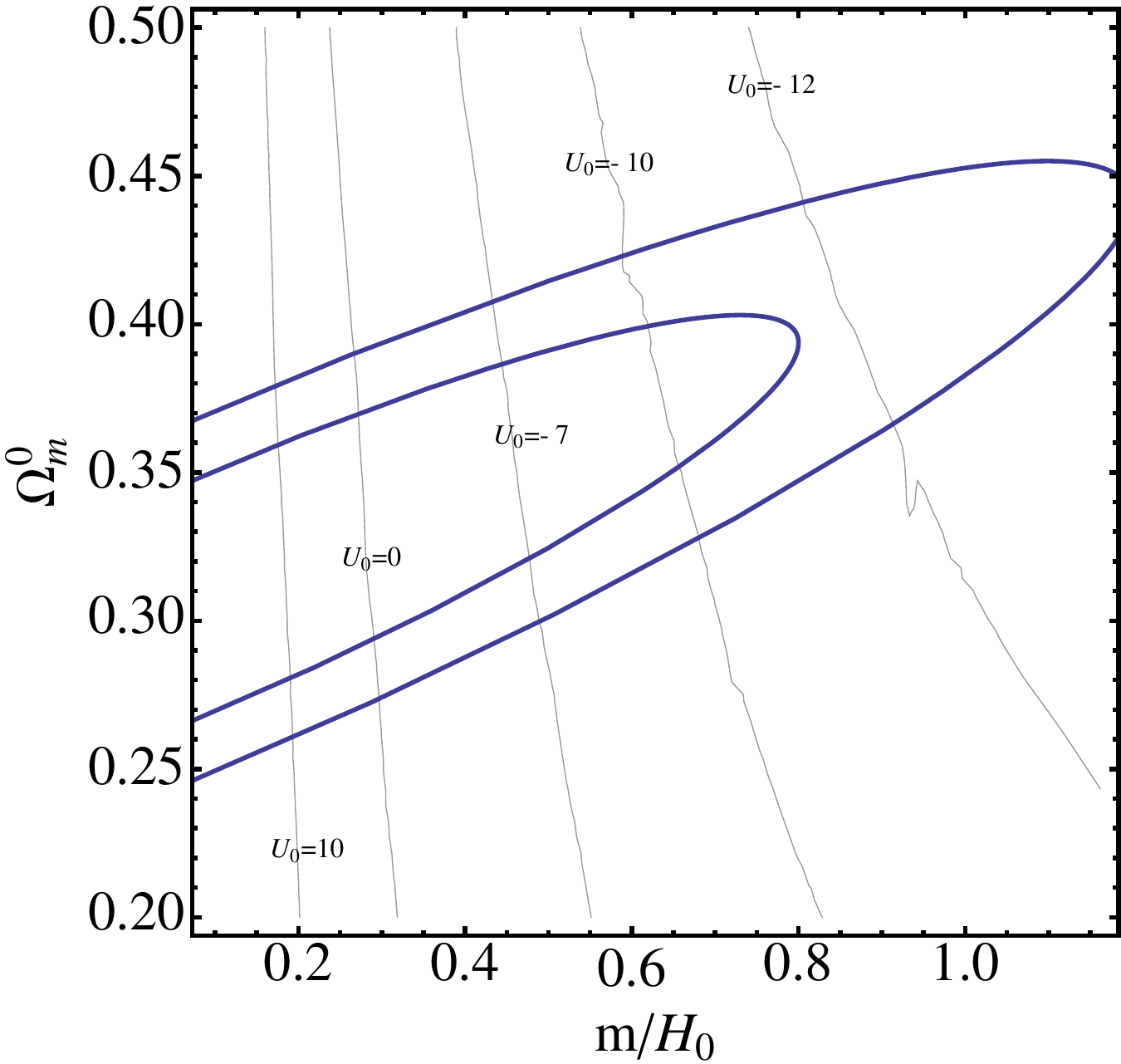} 
\hspace{0.8cm}\includegraphics[height=6cm]{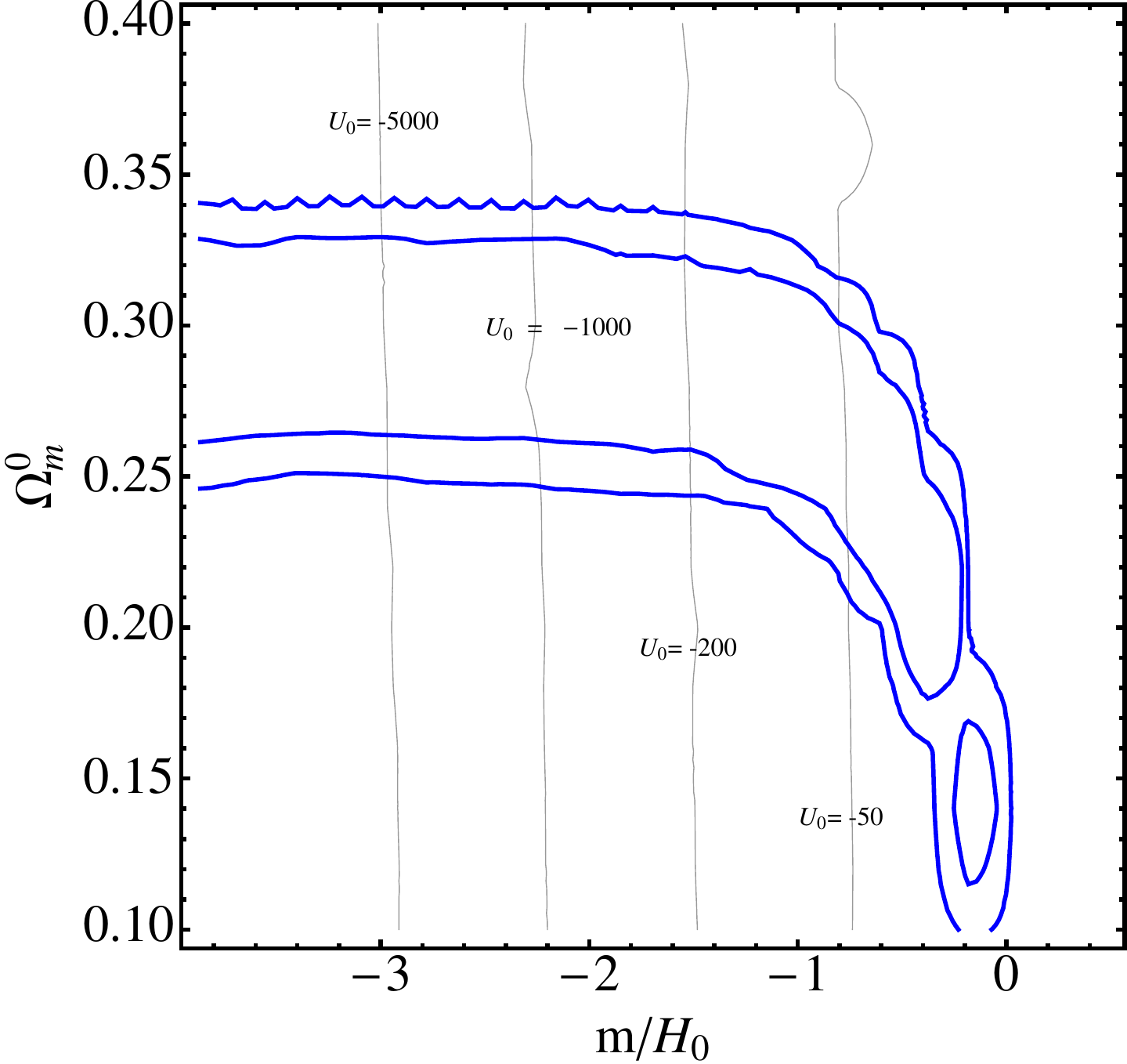} 
\par\end{centering}

\protect\protect\protect\protect\protect\caption{\label{fig:SN-likelihood-contoursU0} Supernovae likelihood contours at 2-$\sigma$ level for path A (left panel) and path B (right panel). The associated values of $U_0$ are also displayed.}
\end{figure}

\begin{figure}
\begin{centering}
\includegraphics[height=6cm]{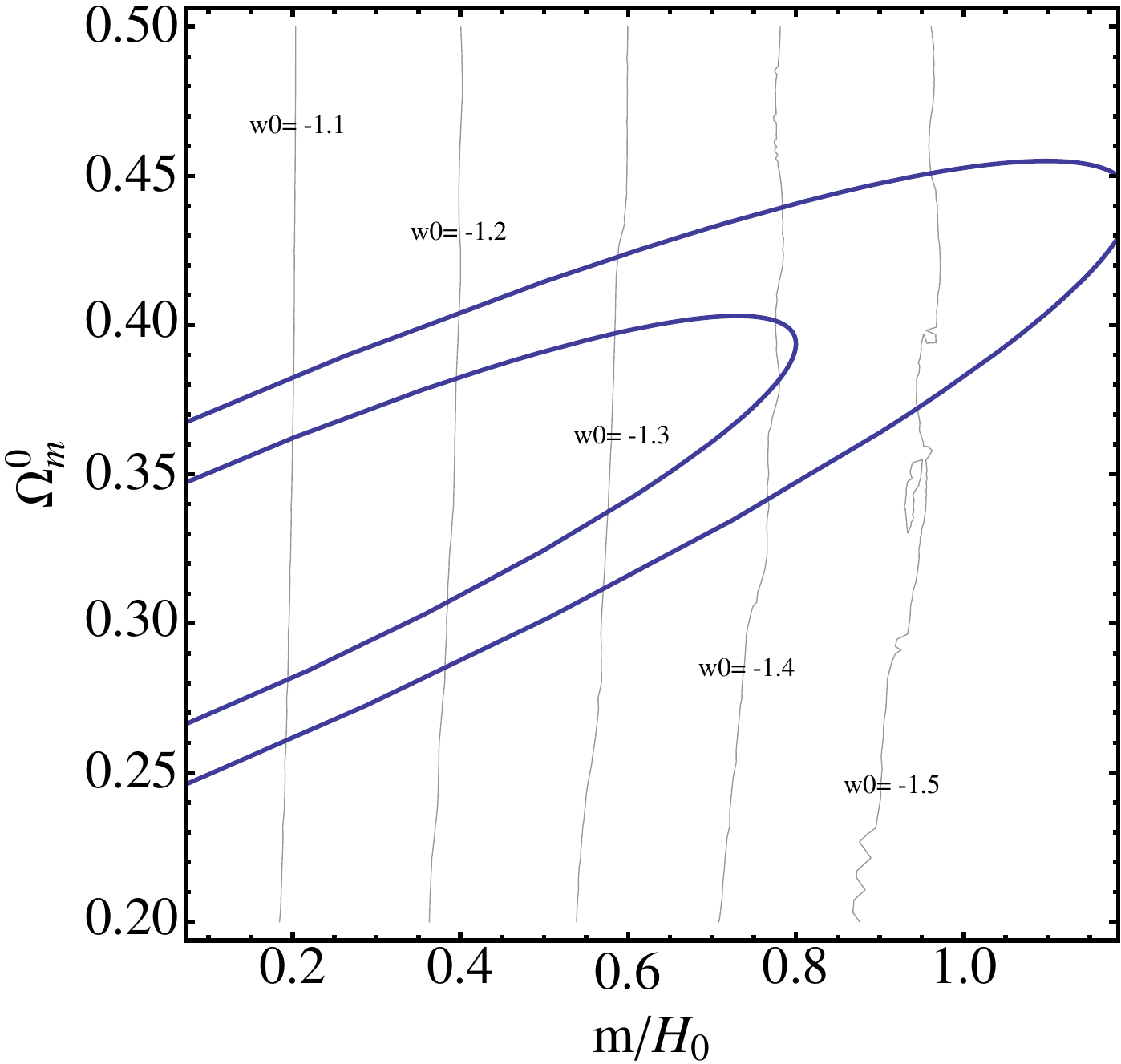} \hspace{0.8cm}\includegraphics[height=6cm]{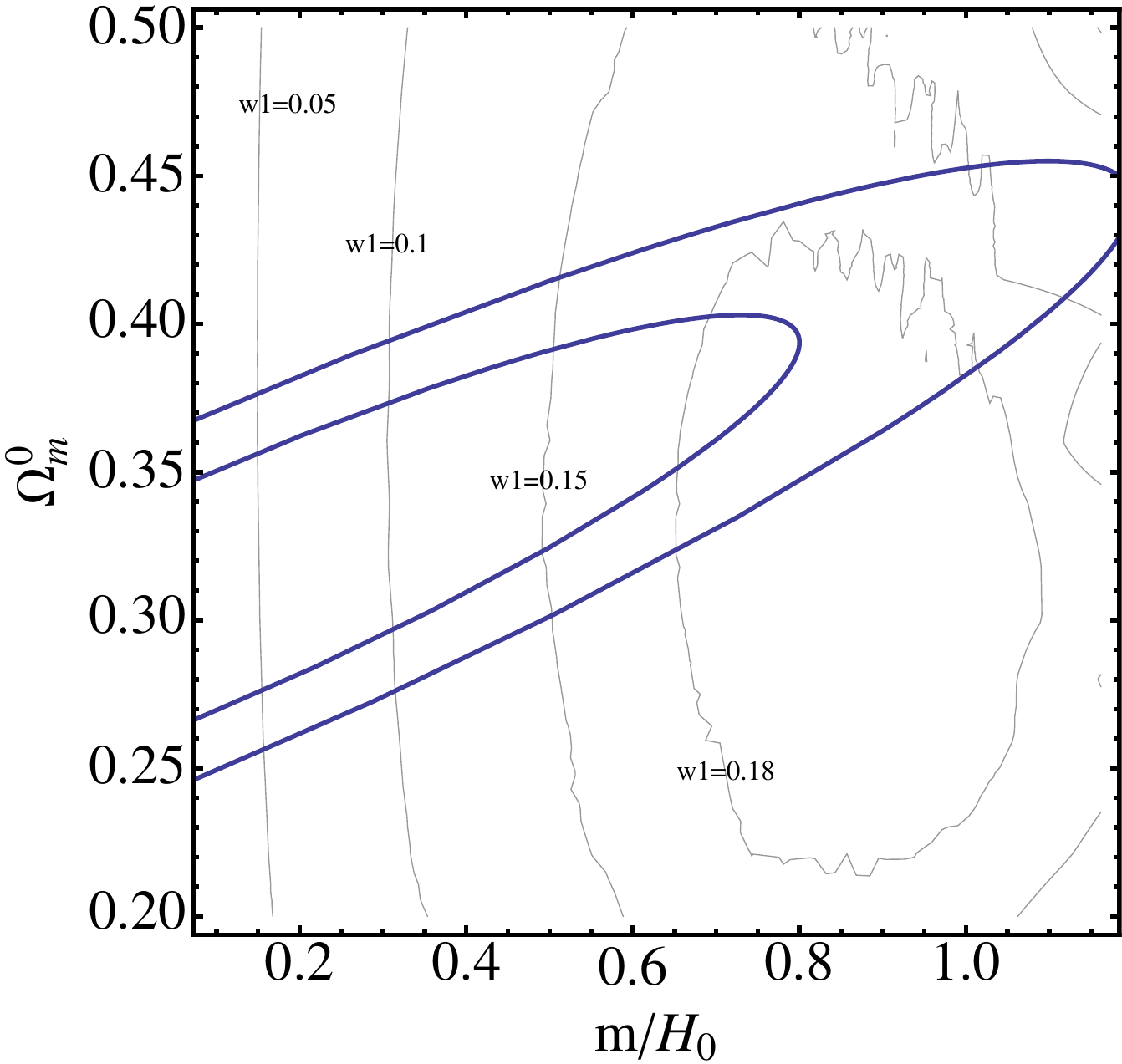} 
\includegraphics[height=6cm]{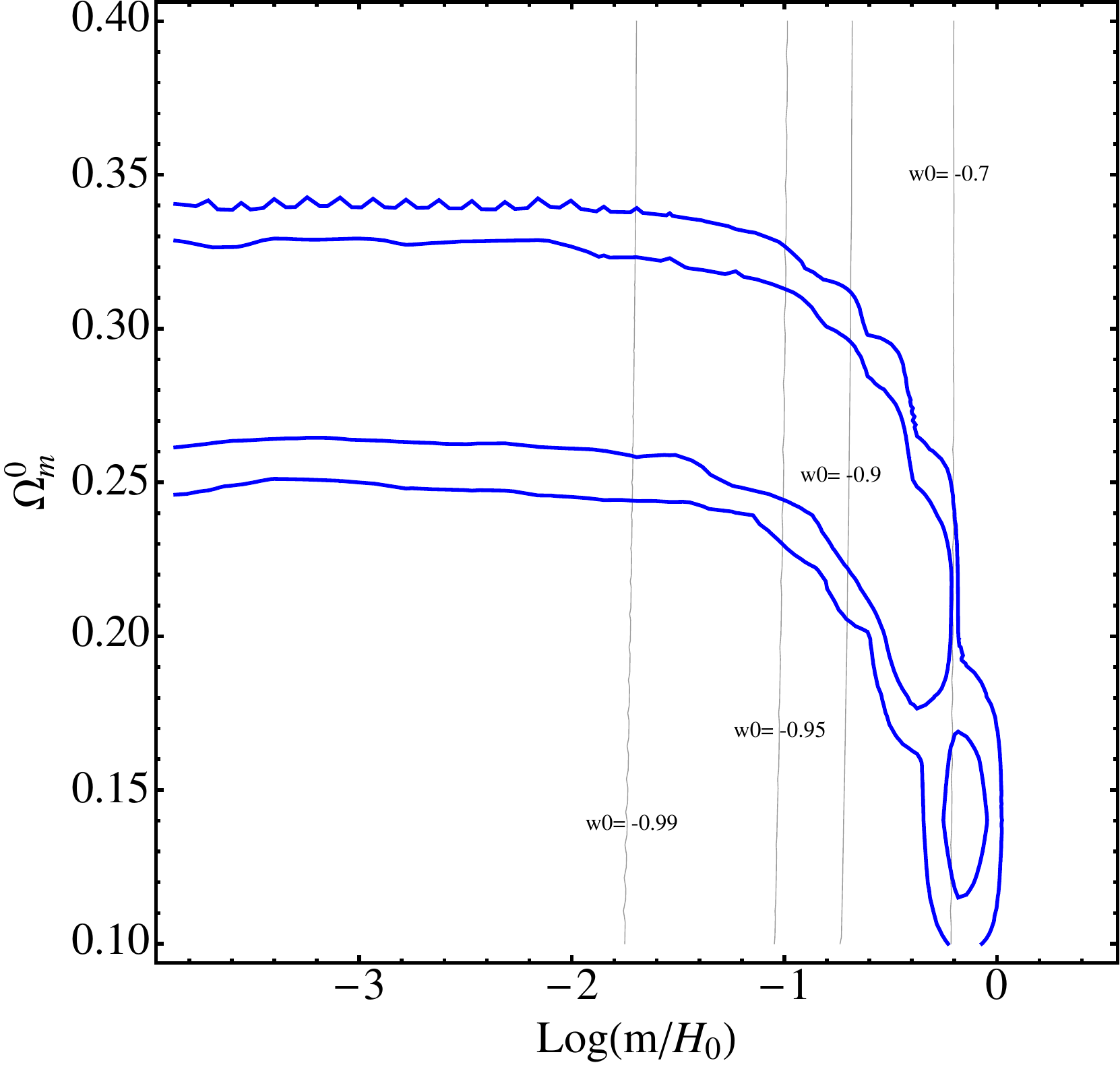} \hspace{0.8cm}\includegraphics[height=6cm]{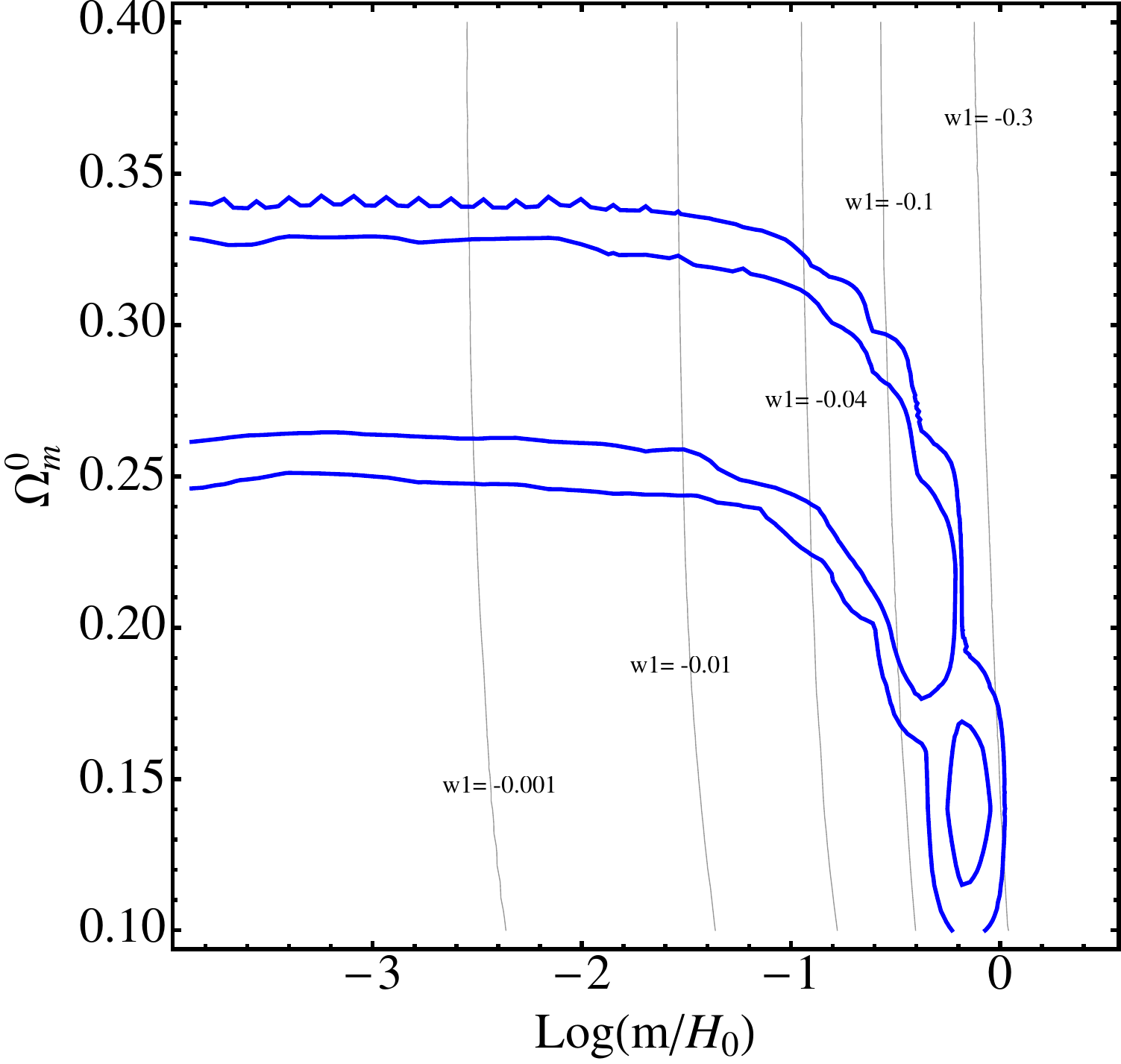} 
\par\end{centering}

\protect\protect\protect\protect\protect\caption{\label{fig:SN-likelihood-contours} Supernovae 
likelihood contours at 2-$\sigma$ level for path A (upper panels) and path B (lower panels). The associated values of 
 $w_0$ and $w_1$ in the standard parametrization $w_{\rm eff}=w_0+(1-a)w_1$ are also displayed.}
\end{figure}

Once the two trajectories are found, we evaluate the Hubble rate $H(z)$
for each path and compare the associated luminosity distance $d_{L}(H(z))$
to the Joint Light-curve Analysis (JLA) supernovae data set \cite{Betoule:2014frx} in order to
obtain two independent likelihoods over $m$ and $\Omega_{\text{M}}^{0}$,
one for each path. When the pair $\lbrace m,\Omega_{\rm M}^0\rbrace$
   is specified, the effective equation of state $w_{\rm eff}$ is completely determined. The results are shown in 
Figs.~\ref{fig:SN-likelihood-contoursU0} and \ref{fig:SN-likelihood-contours}. Focusing on $\Omega_{M}^{0}\approx0.3$, one sees that all
the values of $m$ up to 0.5 are roughly compatible with supernovae.
Note, however, that the expectation value $\Omega_{\text{M}}^{0}\approx 0.3$ 
comes from standard cosmology and it should not be directly applied to modified gravity cases. In fact, the supernovae
data set is roughly compatible with all values of $\Omega_{\text{M}}^{0}<0.45$, so a more robust upper limit for $m$ is around 1.2.
For very small $m$, the trajectories of both path A and path B become observationally
indistinguishable from $\Lambda$CDM.\footnote{Indeed, when $m$ is small, the dynamical part associated
with nonlocal contributions in Eq.~\eqref{eq:h2ithomega} is suppressed. The leading contribution at early 
times is of order $m U_{\rm hom}$,  which is a constant in our case. Note that this is in agreement with the curve 
corresponding to path B in Fig.~\ref{fig:Value-of-}.} 
Note, however, that this may change in the future  when the dynamical part associated
with nonlocal contributions in Eq.~\eqref{eq:h2ithomega} becomes dominant again.

\section{Summary and conclusions}

\label{sec:conclusions}

Nonlocality can emerge from local theories. If one focuses
on classical physics at long wavelengths, there can appear nonlocal constraints due to the effect
of short wavelengths that have been integrated out. In quantum field theories, nonlocality is
introduced in the computation of the effective action via the integration of the radiative corrections 
due to massless or light particles.

It is thus natural to consider that nonlocal, infrared 
modifications of gravity at cosmological scales, such as in the model described by (\ref{NLaction1}),
could provide a useful effective approach to study the problems of cosmological constant and of dark energy.
An important subtlety, not arising in local modifications of gravity, has to then be taken into account in such studies:
in addition to the mass parameter in the Lagrangian (\ref{NLaction1}), the nonlocal model
is understood to be specified by the boundary conditions implied by the presence of the inverse-d'Alembertian. 

In this work, we have considered the effect of general initial conditions
on the dynamical system (\ref{eq:1set})-(\ref{eq:lastset4}) for
the background evolution of the $R\,\Box^{-2}R$ cosmological model,
as well as the constraints from supernovae data on the parameters
$m$ and $\Omega_{\text{M}}^{0}$. The system exhibits two distinct
classes of late-time behavior, which lead to two different types of
cosmological evolution, dubbed path A and path B.

Path A (path B) is realized above (below) a certain threshold $\bar{U}$
for the initial condition of the auxiliary field $U$, $U_{0}$. The
case $U_{0}=0$, belonging to path A, is the one already discussed
in Ref.~\cite{Dirian:2014bma}.  Note, however, that the initial conditions in this theory are sensitive to
the thermal history of the Universe. In particular, the 
value of $U$ at a given number of $e$-foldings cannot be unambiguously set to zero by  
setting $U_{{\rm hom}}=0$. The main result of this work is to
extend the cosmological analysis to the full range of initial conditions.

We found that although both paths possess well-behaved radiation and
matter eras, the subsequent evolution is in general radically different.
Along path A, the system goes through a phantom regime and finally
reaches an attractor on which the effective equation of state $w_{\text{eff}}$
remains frozen at the CC value $-1$. This final state, however, is
not a de Sitter stage since $H$ (and therefore the Ricci scalar $R$)
is not constant, but rather grows indefinitely. Along path B, instead,
the evolution remains always nonphantom; the system reaches generically
a $w_{\text{eff}}=-1$ stage which approaches a true de Sitter stage.
This is however a temporary stage in cosmic evolution, as the solution is not an attractor but, 
rather, a saddle point. After a transient period the system reaches a final configuration
represented by a decelerated, radiationlike, $w_{\text{eff}}=1/3$
state (and therefore with $R=0$), in which, however, no radiation
is present. The present value of the nonlocal-term equation of state,
$w_{\text{DE}}$, can take essentially any values around $-1$. The
impact of initial conditions on the final evolution is therefore important.

Both paths are in principle cosmologically viable. When compared to
supernovae observations, we find the regions in the $\{m,\Omega_{\text{M}}\}$
parameter space that satisfy observational constraints. It is interesting
to note that small, even vanishing, values of $m$ are perfectly acceptable.
This means that cosmologically viable nonlocal terms can be generated
from standard loop corrections, which require $m/H_{0}\ll1$ (in Planck
units). However, in this case the evolution becomes indistinguishable
from $\Lambda$CDM.

The methods employed in this paper can be applied to more
general nonlocal models. We saw that the possible cosmological dynamics
of a given model can be conveniently derived from the behavior of the additional
scalar modes carried by the nonlocal integral operators. Assuming a background evolution
with a power-law expansion of the scale factor, we solved for the mode functions in the
model (\ref{NLaction1}), and from the general solutions (\ref{eq:solu}) and (\ref{eq:solv}) deduced 
the handful of fixed points and their basic properties. A similar analysis should be even more
transparent for actions of the type (\ref{DWaction}), since there the nontrivial modes are
given solely by Eq. (\ref{eq:solu}). In models featuring the conformal Weyl 
curvature \cite{Frank,Cusin:2015rex,Cusin:2016nzi}, a substantial simplification is that only the
homogeneous modes are nonvanishing.

In conclusion, the cosmological background dynamics of the nonlocal model studied here depend
qualitatively upon the initial conditions.  We discussed the initial conditions in terms
of $U_0$ set at an early radiation-dominated epoch and showed that different values for this parameter can result
in the Universe ending up eventually in drastically different stages: in the two most typical cases studied here, 
either a phantomlike approach towards asymptotic singularity, or an eternal conformal expansion. The initial value $U_0=0$
is a natural choice, but its implementation in fact depends on the thermal history of the Universe. As shown
in Appendix \ref{appendix2}, however, the ambiguity is not large enough to change the evolution from the former track
to the latter. 

We can thus regard the interesting dark energy behavior as a robust background prediction of the model (\ref{NLaction1}).
It remains to be seen whether one can learn further about the structure formation in nonlocal cosmology by revisiting the boundary 
conditions of the integral operators at the level of the inhomogeneous perturbation modes.

\acknowledgments We thank Florian F\"{u}hrer and Adam R. Solomon for helpful
discussions. We also acknowledge support from DFG through the project
TRR33 ``The Dark Universe.'' H.N. also acknowledges financial support from DAAD through the 
program ``Forschungsstipendium f\"{u}r Doktoranden und Nachwuchswissenschaftler.''

\appendix

\section{Some clarifications on fixed points and paths}

\label{appendix1}

\subsection{From fixed surfaces to fixed points}

As explained in Sec. \ref{sec:PS_dyn}, when the first derivative
of a variable on the left-hand side of Eqs.~\eqref{eq:1set}-\eqref{eq:lastset4}
takes a constant value, the dynamical system contains a fixed surface
rather than a fixed point. To illustrate how to deal with this situation,
we present below the complete phase-space analysis for the critical
point II in Table \ref{tab:fixedpoints}. A similar analysis can be
done for points III and IV.

As follows directly from Eq.~(\ref{eq:1set}), when $Y_{1}=2$ the
variable $U$ satisfies the equation of a line, $U'=2$. In order
to go from this fixed line to a fixed point we can consider a field
redefinition, 
\begin{equation}
U=\tilde{U}+2N,\label{eq:redefu}
\end{equation}
with $N$ the number of $e$-foldings. Inserting this relation into
Eqs.~\eqref{eq:1set}-\eqref{eq:lastset4}, one immediately realizes
that $Y_{1}=2$ corresponds to a fixed point $\tilde{U}'=0$. Written
in the new variables, the analysis proceeds along the lines discussed
in Sec. \ref{sec:PS_dyn}. The behavior of the system around the
fixed point is determined by the eigenvalues of the characterizing
matrix, which are given by 
\begin{equation}
\lambda_{i}=\left\{ 0,0,-1,-\dfrac{3}{2},-\dfrac{3}{2},\dfrac{3\tilde{U}_{0}+6N+4}{\tilde{U}_{0}+2N}\right\} ,\label{eigen1}
\end{equation}
with $\tilde{U}_{0}$ an arbitrary constant. From the theory of dynamical
systems we know that the so-called Lyapunov coefficients $s_{i}$
($i=1,2,...,6$) are equal to the real part of the eigenvalues $\lambda_{i}$,
provided that these eigenvalues are constant. Note however that, due
to the field redefinition \eqref{eq:redefu}, the last eigenvalue
in Eq.~\eqref{eigen1} depends on the number of $e$-foldings $N$.
The Lyapunov coefficient in this case is defined by the upper limit
\begin{equation}
s_{6}=\lim_{N\rightarrow\infty}\dfrac{1}{N-N_{0}}\int_{N_{0}}^{N}\text{Re}\{\lambda_{6}\left(N'\right)\}\text{d}N',\label{eq:lyapunovcoef}
\end{equation}
with $N_{0}$ some initial value for $N$. Taking into account \eqref{eigen1},
we get 
\begin{align}
s_{6} & =\lim_{N\rightarrow\infty}\dfrac{1}{N-N_{0}}\int_{N_{0}}^{N}\dfrac{3\tilde{U}_{0}+6N'+4}{\tilde{U}_{0}+2N'}\text{d}N'=3.
\end{align}
The resulting spectrum of Lyapunov coefficients 
\begin{equation}
s_{i}=\left\{ 0,0,-1,-\dfrac{3}{2},-\dfrac{3}{2},3\right\} 
\end{equation}
shows that the fixed point under consideration is a saddle point.

\subsection{On the two realizations of the fixed point V}

Note that point V in Table \ref{tab:fixedpoints} has two different
realizations. The first one is obtained for $V=+\infty$ and $\Omega_{\text{M}}=-\infty$,
while the second case corresponds to $V=-\infty$ and $\Omega_{\text{M}}=+\infty$.
In this Appendix, we discuss the set of initial conditions giving
rise to each one of these configurations.

Around the fixed point V, we have $U'=U''=0$. These two conditions
restrict the $\xi$ parameter in Eq.~\eqref{eq:u1} to a fixed value
$\xi=-2$. Inserting this constant value into Eq.~\eqref{eq:v1}
and taking into account the large $N$ limit of Eq.~\eqref{eq:h2ithomega},
we get 
\begin{equation}
V''-3V'-4V+\frac{4}{3\gamma}=0.\label{eq:eqsimplv}
\end{equation}
For $N\gg1$, the solution of this differential equation reads 
\begin{equation}
V\approx\frac{1}{3\gamma}+\left(V_{0}-\frac{1}{3\gamma}+V_{0}'\right)e^{4\left(N-N_{0}\right)},\label{eq:Vexpresnbig}
\end{equation}
with $V_{0}$ and $V'_{0}$ the values of $V$ and $V'$ at some initial
time $N_{0}$. In view of this solution, we can distinguish two possibilities. If
$V_{0}>1/(3\gamma)-V_{0}'$, the system approaches the fixed point
V with $V\rightarrow+\infty$.%
\footnote{ Note that realizations with $V_{0}>1/(3\gamma)-V_{0}'$ are not physically
acceptable. As can be easily deduced from Eq.~(\ref{eq:h2ithomega}),
these configurations give rise to negative values of $h^{2}$ around
the fixed point $V$.%
} In the opposite case, the fixed point V is realized with $V\rightarrow-\infty$.

\section{Initial conditions and thermal history of the Universe}

\label{appendix2}

In this Appendix we estimate the robustness of the MM initial conditions
when the detailed particle content of the Universe prior to matter-radiation
equality is taken into account. 

For some purposes, the transition from radiation to matter domination
can be approximated by an instant transition at $N_{\text{eq}}\simeq-8.1$
in which the trace of the energy-momentum tensor changes abruptly
from zero to $e^{3N_{\text{eq}}}$ times its present value. This approximation
implicitly assumes that all the particles in the early Universe have
roughly the same mass, and transit simultaneously from a relativistic
to a nonrelativistic state. A detailed analysis of the thermal history
of the Universe allows us to go beyond this approximation and to account
for the fact that particles with different masses become nonrelativistic
at different temperatures, or equivalently, at different cosmic times.

The change in the trace of the energy-momentum tensor can be parametrized
as 
\begin{equation}
\mathrm{tr}\left(T_{\mu\nu}\right)=\frac{\rho-3P}{\Mp^{2}}\equiv\frac{\rho_{\text{R}}}{\Mp^{2}}\Sigma\left(T\right),\label{eq:rkick}
\end{equation}
with $T$ and $\rho_{\text{R}}$ the temperature and energy density
of the radiation bath.%
\footnote{Note that if all the particles prior to recombination were completely
massless, $\Sigma(T)$ would be zero.%
} The so-called ``kick'' function $\Sigma$(T) is computed by summing
the individual contributions to the bath of the particles with mass
$m_{i}$, temperature $T_{i}$ and $g_{i}$ degrees of freedom \cite{Erickcek:2013dea},%
\footnote{The $+$ and $-$ signs in the denominator of the integrand apply,
respectively, to fermions and bosons. %
} 
\begin{equation}
\Sigma_{i}\left(T\right)=\frac{\rho_{i}-3p_{i}}{\rho_{i}}=\frac{15}{\pi^{4}}\frac{g_{i}}{g_{\ast}\left(T\right)}\left(\frac{m_{i}}{T}\right)^{2}\int_{m_{i}/T}^{\infty}\frac{\sqrt{u^{2}-\left(m_{i}/T\right)^{2}}}{e^{u}\pm1}\text{d}u\,,\label{eq:sigmaind}
\end{equation}
where $g_{\ast}\left(T\right)\equiv\rho_{\text{R}}[(\pi^{2}/30)T^{4}]^{-1}$
is the total number of relativistic degrees of freedom in the bath.

Equation~\eqref{eq:rkick} translates, via Einstein equations, into a
change of the Ricci scalar, $R=\mathrm{tr}\left(T_{\mu\nu}\right)$.
Integrating Eq.~\eqref{eq:udefin} with $a(t)\simeq t^{1/2}$ and
taking into account the relation between time and temperature at radiation-domination,
\begin{equation}
t\simeq\sqrt{\frac{45}{2\pi^{2}}}g_{\ast}^{-1/2}\frac{\Mp}{T^{2}},
\end{equation}
we get 
\begin{align}
\Delta U & \equiv U-U_{\text{hom}}=\frac{45}{2\pi^{2}}\Mp^{2}\int_{T_{\text{i}}}^{T_{\text{f}}}\text{d}T'{\cal D}(T')g_{\ast}^{1/4}\left(T'\right)T'\times\int_{T_{\text{i}}}^{T'}\text{d}T''{\cal D}(T'')\frac{R\left(T''\right)}{\left(T''\right)^{5}g_{\ast}^{5/4}\left(T'\right)},\label{eq:usigma}
\end{align}
with 
\begin{equation}
{\cal D}(T)\equiv\left(-\frac{2}{T}-\frac{1}{2g_{\ast}\left(T\right)}\frac{\partial g_{\ast}\left(T\right)}{\partial T}\right).
\end{equation}
Here, $T_{\text{i}}$ and $T_{\text{f}}$ are the higher and lower
temperatures for which radiation domination is a reasonable first
order approximation for the background evolution of the Universe.
Note that, due to the integration between $T_{\text{i}}$ and $T_{\text{f}}$,
even a tiny value of the scalar curvature at early times ($T_{\text{f}}\ll T_{\text{i}}$)
can give rise to a sizable modification of $U$ at matter-radiation
equality.%
\footnote{For $T_{\text{i}}\gg T_{\text{f}}$ and constant values of $g_{*}$
and $\Sigma$ we have 
\begin{align*}
\Delta U\sim\,\Sigma\int_{T_{\text{i}}}^{T_{\text{f}}}\text{d}T'\int_{T_{\text{i}}}^{T'}\text{d}T''\frac{1}{\left(T''\right)^{2}}\sim\Sigma\left[\ln\frac{T_{\text{i}}}{T_{\text{f}}}-\frac{T_{\text{i}}-T_{\text{f}}}{T_{\text{i}}}\right]\sim\Sigma\ln\frac{T_{\text{i}}}{T_{\text{f}}}.
\end{align*}
}

The radiation-domination requirement ($R=0$) giving rise to the MM initial conditions
should be understood only as an approximation of the actual dynamics. The value of $U$ at radiation domination cannot be unambiguously
set to zero by simply setting $U_{{\rm hom}}=0$. Indeed, if the initial conditions are set at the end of inflation/reheating,
one should expect nonvanishing values of $U_{0}$ at the number of 
$e$-foldings at which the MM initial conditions are usually implemented ($N\simeq -14$). Note also that, even if the MM initial 
conditions are taken for granted, the detailed thermal history of the Universe will inevitably affect the subsequent evolution 
of $U$. The uncertainty associated to this effect depends on the particle content of the early Universe. Assuming 
radiation domination between $1000$ GeV and $0.75$ eV and considering only the contribution of Standard
Model particles, we can numerically integrate Eq.~\eqref{eq:usigma}
to obtain a correction
\begin{equation}
\Delta U\approx1.6\,,\label{Umin}
\end{equation}
to be added on top of the nonvanishing MM value at matter-radiation equality~\cite{Dirian:2014ara}.  The assumptions leading to the 
uncertainty~\eqref{Umin}  are indeed quite conservative. Larger
values of $\Delta U$ should be expected if we accept the existence
of new physics beyond the Standard Model. 

\bibliographystyle{apsrev}
\bibliography{amendolamodnonloc,Henrik,nonlocalRefsTSK}

\end{document}